\documentclass[12pt]{article}

\usepackage{amssymb,amsmath,color,graphicx,dsfont,mathrsfs,amsthm,stmaryrd}
\usepackage{authblk}
\usepackage{faktor}

\usepackage{float}

\newtheorem{theorem}{Theorem}
\newtheorem{corollary}[theorem]{Corollary}
\newtheorem{lemma}[theorem]{Lemma}
\newtheorem{proposition}[theorem]{Proposition}
\newtheorem{definition}[theorem]{Definition}
\newtheorem{remark}[theorem]{Remark}

\renewcommand{\th}{\theta}

\newcommand{\bt}{\begin{theorem}}
\newcommand{\et}{\end{theorem}}
\newcommand{\bl}{\begin{lemma}}
\newcommand{\el}{\end{lemma}}
\newcommand{\bp}{\begin{proposition}}
\newcommand{\ep}{\end{proposition}}
\newcommand{\bc}{\begin{corollary}}
\newcommand{\ec}{\end{corollary}}
\newcommand{\bdeff}{\begin{definition}}
\newcommand{\edeff}{\end{definition}}
\newcommand{\brem}{\begin{remark}}
\newcommand{\erem}{\end{remark}}
\newcommand{\bproof}{\begin{proof}}
\newcommand{\eproof}{\end{proof}}

\newcommand{\bi}{\begin{itemize}}

\renewcommand{\i}{\item}
\newcommand{\ei}{\end{itemize}}
\newcommand{\bd}{\begin{description}}
\newcommand{\ed}{\end{description}}
\newcommand{\be}{\begin{enumerate}}
\newcommand{\ee}{\end{enumerate}}

\newcommand{\bqn}{\begin{eqnarray}}
\newcommand{\eqn}{\end{eqnarray}}
\newcommand{\eqnn}{\nonumber\end{eqnarray}}
\newcommand{\eqnl}[1]{\end{eqnarray}}

\newcommand{\ba}[1]{\begin{array}{#1}}
\newcommand{\ea}{\end{array}}

\newcommand{\R}{\mathbb{R}}
\newcommand{\Z}{\mathbb{Z}}
\newcommand{\N}{\mathbb{N}}
\newcommand{\C}{\mathbb{C}}

\newcommand{\F}{\mathcal{F}}
\newcommand{\D}{\mathcal{D}}

\newcommand{\al}{\alpha}
\newcommand{\eps}{\varepsilon}

\newcommand{\Tf}{\frac{1}{\eps_1 \eps_2}}

\title{Ensemble qubit controllability with a single control via adiabatic and rotating wave approximations} 
\author[1]{R\'emi Robin\thanks{remi.robin@inria.fr}$^,$ }
\author[2]{Nicolas Augier}
\author[1]{Ugo Boscain}
\author[1]{Mario Sigalotti}

\affil[1]{\footnotesize Laboratoire Jacques-Louis Lions, Sorbonne Universit\'e, Universit\'e de Paris, Inria, CNRS, France}
\affil[2]{\footnotesize Inria, Universit\'e C\^ote d'Azur, INRAE, CNRS,  Sorbonne Universit\'e, Sophia Antipolis, France}

\begin{document}

\maketitle

\begin{abstract}

In the physics literature it is 
common to see the  rotating wave approximation and the adiabatic approximation used ``in cascade'' to justify the use of chirped pulses for two-level quantum systems driven by one external field, in particular when the resonance frequency of the system is not known precisely.

Both approximations need relatively long time and are essentially based on averaging theory of dynamical systems.
Unfortunately, the two approximations cannot be done independently since, in a sense, the two time scales interact. 
The purpose of this paper is to study how the cascade of the two approximations can be justified 
and how large becomes the final time as the fidelity goes to one, while preserving the robustness of the adiabatic strategy.
Our first result, based on high-order averaging techniques, gives a precise quantification of the uncertainty interval of the resonance frequency for which the population inversion works.
As a byproduct of this result, we prove that it is possible to control an ensemble of spin systems by a single real-valued control, providing a non-trivial extension of a celebrated result of ensemble controllability with two controls by Khaneja and Li.
 
\end{abstract}



\section{Introduction}


Consider a two-level system described by the
Schr\"odinger equation 
\begin{equation}
\label{1}
i \frac{d\psi}{dt}=
\Big(
\begin{array}{cc}
E+\al& w(t)\\w(t)&-E-\al \end{array}
\Big)\psi.
\end{equation}
Here $w:[0,T]\to\R$ is a (sufficiently regular) function representing an external field, 
$E>0$, and $\al\in[\al_0,\al_1]$ is an unknown parameter representing the fact that the {\em resonance frequency} of the system  $2(E+\al)$ is not known precisely, but lies between $2(E+\al_0)$ and $2(E+\al_1)$.
All along the paper we assume the condition 
 \begin{equation*}
[\al_0,\al_1]\subset (-E,\infty),\qquad 0\in(\al_0,\al_1),
\end{equation*}
guaranteeing that the eigenvalues of the matrix in equation \eqref{1} are never zero, independently of the value of $\alpha$.  
The solution of  \eqref{1}  (that depends on $\alpha$ and $w(\cdot)$) with initial condition $\psi^\al_w(0)=(0,1)$ is the wave function $\psi_w^\al:[0,T]\to\C^2$.  

One would like to find a function $w(\cdot)$ (the same for all values of $\al$) such that, if at time zero the system is at the ground state $(0,1)$ (i.e., it is in the eigenstate corresponding to the eigenvalue $-E-\al$), then at time $T$ the system is close to a state of the form $(e^{i\th},0)$
for some $\th\in\R$. In mathematical terms this can be rephrased as follows.

\medskip
{\bf P}: {\em For  every $\eps>0$, 
find a time $T$ and an 
external field $w:[0,T]\to\R$ 
such that 
$$
|\psi^\al_w(T)- (e^{i\th},0)|< \eps, 
$$
for every $\alpha\in[\al_0,\al_1]$ and for some $\th\in\R$ (possibly depending on $\eps,E,T,w,\al$).}\\[2mm]
In the mathematical literature it has been proved that problem  {\bf P} admits a solution when one replaces the real-valued function $w$ by a complex-valued one, as in equation~\eqref{1rwa} below (\cite{BCR,k1,k2,cdc-dasilva}). 
As far as we are aware, the problem is open in the case of real-valued functions.
The result proved in this paper (Theorem \ref{mainTh}) solves problem {\bf P} in a more general framework, in which there is an additional parameter dispersion on the coupling between the control and the system (that is, $w(t)$ is replaced by $\delta w(t)$ for $\delta$ in a compact interval of $(0,+\infty)$).

Solving {\bf P} is a key ingredient to prove 
ensemble controllability of \eqref{1} with more general initial and final conditions.
This celebrated problem has been solved in the case where $w$ is replaced by a complex-valued control in \cite{k2,k1} and \cite{BCR}.

The intuitive approach to tackle problem {\bf P}, consists in the following two steps (\cite{mittleman2013introduction,shore-article,shore_2011,Vitanov-article}):
\begin{itemize}
\item use an external field oscillating at the resonance frequency $2E$ and having a small and slowly varying amplitude and a slowly varying phase, to simulate by rotating wave approximation (RWA, for short) a system driven by a complex-valued function (in a sense, this ``duplicates'' the number of available external fields);
\item use an adiabatic strategy based on chirped pulses (i.e., pulses whose frequency is slowly increasing from a value below $2(E+\al_0)$ to a value above $2(E+\al_1)$) to drive the system from an eigenstate to the other one independently of the value of $\alpha$. This second step substantially exploits the presence of a complex-valued external field and is called adiabatic approximation (AA, for short) \cite{Garwood,Malinovsky,ccch11,PhysRevLett.106.166801,PhysRevLett.106.067401}. Alternative robust methods are developed, for example, in \cite{Jo2017,PhysRevLett.106.233001}.
\end{itemize}
However the RWA may affect the precision of the adiabatic strategy, as it has been remarked in \cite{rouchon-sarlette}.
In order to detail in which sense the ``cascade'' of  the two approximations 
introduced above may
break down, let us give some quantitative  estimate.

\subsection{Rotating wave approximation}  \label{sec:RWA}
Consider a two-level system of the form 
\begin{equation}
i \frac{d\psi}{dt}=
\Big(
\begin{array}{cc}
E& w(t)\\w^\ast(t)&-E\end{array}
\Big)\psi.
\label{1rwa}
\end{equation}
Here we assume that the resonance frequency of the system is known precisely, hence we have no $\al$. The symbol $w^\ast$ denotes the complex conjugate of  $w$, which represents here a complex-valued external field.
For every $\eps>0$, consider the external fields
\begin{align}
w_\eps(t)&= 2 \eps  u(\eps t) \cos(2 E t +\Delta(\eps t)),
\label{contr-rwa1}\\
{w_\eps^{\mathrm{R}}}(t)&= \eps  u(\eps t) e^{-i(2 E t +\Delta(\eps t)) }.
\label{contr-rwa2}
\end{align}
where $u(\cdot)$ and $\Delta(\cdot)$ are two real-valued smooth 
functions defined on $[0,T]$, $T>0$.
We have the following.
\begin{proposition}
For $\eps>0$ let $\psi_{w_\eps}$ and $\psi_{w_\eps^{\mathrm{R}}}$ be the solutions  of \eqref{1rwa} with initial condition $\psi_0\in \C^2$ corresponding to the external fields $w_\eps$ 
and ${w_\eps^{\mathrm{R}}}$, respectively.
Then $\max_{t\in [0,T/\eps]}|\psi_{w_\eps}(t)-\psi_{w_\eps^{\mathrm{R}}}(t)|$  converges 
to $0$ as $\eps\to 0$. 
\end{proposition}
The proof of this fact is well known. If one applies the unitary change of variables
\[
\psi_{w_\eps}(t)=\Big(
    \begin{array}{cc}
    e^{-i(E t+\Delta(\eps t)/2)}& 0\\0&e^{i(E t+\Delta(\eps t)/2)}\end{array}
    \Big)
    \hat\psi_{w_\eps}(t)\]
then $ \hat\psi_{w_\eps}(t)$ satisfies the Schr\"odinger equation 
\begin{align*}
i &\frac{d\hat\psi_{w_\eps}}{dt}=
\eps\Big[ \Big(
\begin{array}{cc}
-\Delta'(\eps t)/2&
u(\eps t)\\ 
u(\eps t)&\Delta'(\eps t)/2\end{array}
\Big)+
 \Big(
\begin{array}{cc}
0&
e^{i( 4 E t+2\Delta(\eps t))}u(\eps t)\\ 
e^{-i( 4 E t+ 2\Delta(\eps t))}u(\eps t)&0\end{array}
\Big)\Big]\hat\psi_{w_\eps}.
\end{align*}
Here $\Delta'$  indicates the derivative of the function $\Delta:[0,T]\to\R$. Now, defining $s=\eps\, t$,  varying in the interval $[0,T]$, and $\tilde\psi_{w_\eps}(s)=\hat\psi_{w_\eps}(t/\eps)$  we obtain
\begin{align}
i &\frac{d\tilde\psi_{w_\eps}}{ds}=
\Big[ \Big(
\begin{array}{cc}
-\Delta'( s)/2&
u(s)\\ 
u(s)&\Delta'(s)/2\end{array}
\Big)+ \underbrace{\Big(
\begin{array}{cc}
0&
e^{i( 4 E s/\eps+2\Delta(s))}u(s)\\ 
e^{-i( 4 E s/\eps+ 2\Delta(s))}u(s)&0\end{array}
\Big)}_{=:B(s,\eps)}\Big]\tilde\psi_{w_\eps}.
\label{rwa-a}
\end{align}
The same change of variables on $\psi_{w_\eps^{\mathrm{R}}}$ gives rise to 
\begin{equation}
i \frac{d\tilde\psi_{w_\eps^{\mathrm{R}}}}{ds}=
\Big(
\begin{array}{cc}
-\Delta'( s)/2&
u(s)\\ 
u(s)&\Delta'(s)/2\end{array}
\Big)
\tilde\psi_{w_\eps^{\mathrm{R}}}.
\label{rwa-b}
\end{equation}

Equations \eqref{rwa-a} and \eqref{rwa-b}  differ only for the term $B(s,\eps)$. Since for every interval $[s_1,s_2]\subseteq[0,T]$ we have 
\[\lim_{\eps\to0} \int_{s_1}^{s_2} B(s,\eps)=0\]
and $B$ is uniformly bounded, we have that solutions of \eqref{rwa-a} converge uniformly in $[0,T]$ to solutions  of \eqref{rwa-b} with the same initial condition. This is a classical averaging result that can be found, for instance, in \cite[Chapter 8]{agrachevbook}.
Coming back to the original variables one obtains that  $|\psi_{w_\eps}-\psi_{w_\eps^{\mathrm{R}}}|$ converges uniformly to zero on the interval $[0,T/\eps]$.

This simple argument is very useful. We started with a system driven by one scalar function $w$ and we obtain at the limit a system driven by 
a complex-valued control or, equivalently, system~\eqref{rwa-b}
where the controls are the
two  scalar functions $u(t)$ and $v(t)=\Delta'(t)/2$.
A more detailed quantitative analysis permits to conclude that on $[0,T/\eps]$ we have
$$
|\psi_{w_\eps}-\psi_{w_\eps^R}|=O(\eps).  
$$
(See, for instance, \cite[Appendix A]{semiconical} for a quantitative version of the averaging result mentioned above.) Higher order RWA can be obtained by considering higher-order averaging results.

In recent applications, it is sometimes necessary to use intense external fields. In these cases the RWA may become inaccurate, as pointed out in \cite{PhysRevA.75.063414,PhysRevA.82.022119,Scheuer_2014}. Thus it is crucial to have a precise quantification of the error.

\subsection{Adiabatic approximation} \label{sec:AA}
We have seen in the previous section how to 
make the solutions of system~\eqref{1rwa}
approximate those of system \eqref{1rwa}. 
We show here how such a system  can be easily driven by adiabatic pulses.


Let us consider the case in which the energy of the system is not known precisely. 
We are then considering the system
\begin{equation}\label{ante2}
i \frac{d\psi}{dt}=
\Big(
\begin{array}{cc}
E+\al& w(t)\\w^\ast(t)&-E-\al\end{array}
\Big)\psi,\quad \mbox{ where }  \al\in[\al_1,\al_2].
\end{equation}
Let us choose the pulse $w$ in the form
\begin{equation}
w(t)= u(t) e^{-i(2 E t +\Delta(t)) },
\label{contr-C}
\end{equation}
where 
 $u(\cdot)$ and $\Delta(\cdot)$ are two real-valued smooth 
functions. This choice of control corresponds to 
\eqref{contr-rwa2} in which 
$\eps$ has been set equal to $1$.
Applying the change of variables  
\[\psi(t)=\Big(
    \begin{array}{cc}
    e^{-i(E t+\Delta( t)/2)}& 0\\0&e^{i(E t+\Delta(t)/2)}\end{array}
    \Big)\Psi(t),\]
we obtain 
\begin{equation}
i \frac{d\Psi}{ds}=
 \Big(
\begin{array}{cc}
\al-v(s)&
u(s)\\ 
u(s)&-\al+v(s)\end{array}
\Big)
\Psi.
\label{2}
\end{equation}
where $v(t):=\Delta'(t)/2$. 

\begin{figure}
\begin{center}\includegraphics[scale=.1]{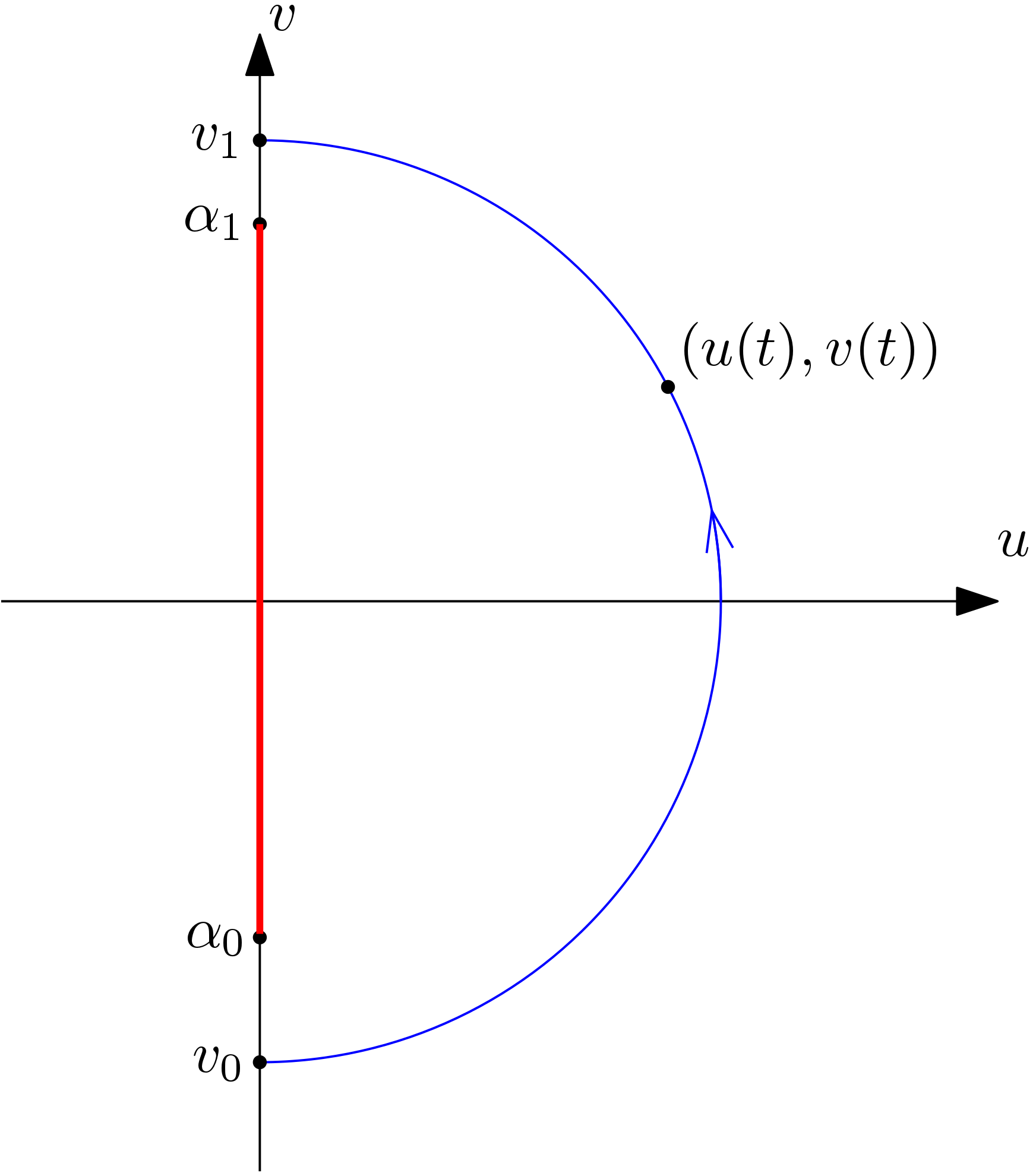}
\caption{An adiabatic path as the one applied in Proposition~\ref{adiabatiqueWithoutRWA}.\label{fig1}}
\end{center}
\end{figure}

Notice that the eigenvalues of the matrix in equation \eqref{2}, seen as functions of the pair $(u,v)$, coincide if and only if  $u=0$ and $v=\al$, where a conical eigenvalue intersection occurs.
Fix now
$v_0<\al_0$ and $v_1>\al_1$ and consider a smooth 
path $t\mapsto(u(t),v(t))$ lying in the half-plane $u>0$ except for the initial and final points, where $u=0$ (see Figure~\ref{fig1}).

Define 
$$
u_\eps(t)=u(\eps t),\qquad v_\eps(t)=v(\eps t).
$$
Since the eigenvalues of the Hamiltonian in Equation \eqref{2} are
\[\pm \sqrt{(\al-v(s))^2+u(s)^2}\ne 0,\] 
 the adiabatic theorem (see, e.g., \cite{teufel-book}) ensures that, for $\eps>0$ small, the trajectory of \eqref{2} corresponding to $(u_\eps,v_\eps)$ and
 starting from $(0,1)$ stays close to the eigenvector associated with the negative eigenvalue.
 More precisely, we have the following estimate.
\begin{proposition}
\label{adiabatiqueWithoutRWA}
There exists $C>0$ such that, for every $\al\in[\al_0,\al_1]$ and every $\eps>0$, the solution $\Psi$ of system
 \eqref{2} with initial condition $(0,1)$ and corresponding to the 
 control $(u_\eps,v_\eps)$ satisfies
$|\Psi(T/\eps)-(e^{i\theta},0)   |\le C\eps$ for some $\theta\in\R$.
\end{proposition}

Going back to equation~\eqref{ante2}, the control 
corresponding to $(u_\eps,v_\eps)$ is 
\[
w_\eps(t)=u(\eps t)e^{i(Et+ \frac{\Delta(\eps t)}{\eps})}.
\]
Such a law  is called a (amplitude modulated) \emph{chirped pulse}, since the range of frequency swept by the pulse is $\{2E+\Delta'(s)\mid s\in [0,T]\}$, which is independent of $\eps$. For more details, see~\cite{MR3874014}.

\subsection{Combination of RWA and AA and statement of the population inversion result}

What one would like to do is to consider the two approximations in cascade, in order to induce a transition from the state $(0,1)$ to $(1,0)$ (up to a phase) for an ensemble of systems parameterized by $\al \in [\al_0,\al_1]$ using a real-valued external field.
The cascade of the two approximations is expected to behave well in many experimental setups, such as in NMR, due to the separation of timescales between the RWA and the AA. However, for intense external fields or in presence of large parametric dispersions, the outcome of the cascade is more challenging to predict and quantify precisely.
Let us denote by $\eps_1$ the small parameter that in the RWA was called $\eps$ and by $\eps_2$ what in the AA was called $\eps$. A formal cascade of the two approaches yields a control law of the form
\[
w_{\eps_1,\eps_2}(t)=2\eps_1 u(\eps_1\eps_2 t) \cos\Big(2 Et+ \frac{\Delta(\eps_1\eps_2 t)}{\eps_2}\Big),
\]
where $u(\cdot)$ and $v(\cdot)$ are the same functions as those used in Proposition~\ref{adiabatiqueWithoutRWA}.

The hope is that the pulse $w_{\eps_1,\eps_2}$, for $\eps_1$ and $\eps_2$ small, induces approximately a transition from the state $(0,1)$ to a state of the form $(e^{i\theta},0)$ in time $T/(\eps_1\eps_2)$.
The two approximations are, however, competing: when one decreases $\eps_2$ (better AA), one needs the RWA to be true for a longer time as the final time is of order $1/(\eps_1\eps_2)$. On the other hand,  decreasing $\eps_1$ deteriorates the performances of the AA:
\begin{enumerate}
\item The error on the adiabatic theorem depends of the gap between the eigenvalues, 
which goes to zero as $\eps_1 \to 0$;
\item The range of frequencies swept by the pulse is $\{2E+ \eps_1\Delta'(s)\mid s\in [0,T]\}$, that is,  the allowed dispersion on the frequency is shrinking as $\eps_1$ goes to zero.
\end{enumerate}
As a consequence, this method can only work when $\al=0$. Under this restriction, and for suitable relations between $\eps_1$ and $\eps_2$ as they both go to  zero, the cascade of the two approximations can be proved to work (see \cite{augier:hal-02277852,RWAEXTENDED}).

\begin{figure}[ht]
\begin{center}
\includegraphics[width=0.7\columnwidth]{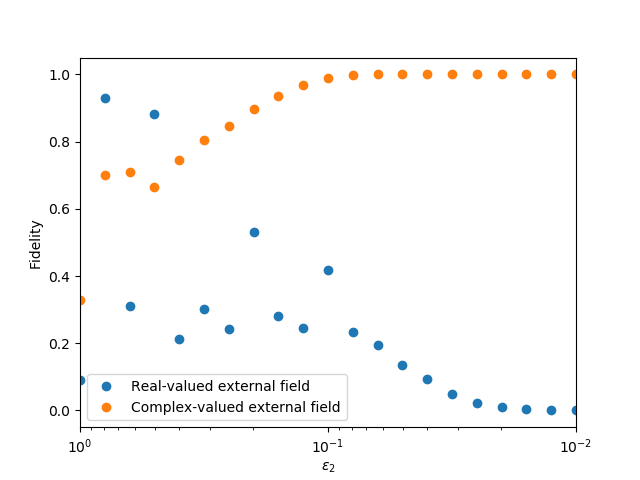}
\caption{Comparison of the real-valued and complex-valued chirp scheme of the first point of Remark~\ref{remark:5} with $E=0.75$, $\al=0.25$, $\eps_1=1$, $v_0=-0.5$, $v_1=0.5$. Notice that the assumptions of Theorem~\ref{mainTh} are not satisfied. \label{fig:outofinterval}}
\end{center}
\end{figure}

Another possibility would be to fix $\eps_1$ small and to hope that the limit as $\eps_2\to0$ makes the RWA  work  as well. Nevertheless, 
the $k$-th order RWA is usually valid up to a time of order $\frac{1}{\eps_1^k}$, whereas we would need the time to be of order $\frac{1}{\eps_1\eps_2}$. In fact, without restriction on the allowed frequency, simulations show that convergence does not hold, as illustrated in Figure~\ref{fig:outofinterval}.

An approach to tackle the issue of the shrinking interval of frequencies swept by the pulse is to divide $\Delta(\eps_1\eps_2 t)$ by $\eps_1\eps_2$ and not just by $\eps_2$.
We claim that an external field of the type
\begin{equation}
\label{control}
w_{\eps_1,\eps_2}(t)=2\eps_1 \delta u(\eps_1\eps_2 t) \cos\Big(2 Et+ \frac{\Delta(\eps_1\eps_2 t)}{\eps_1\eps_2}\Big),
\end{equation}
where $\delta$ is a positive constant, can induce a transition for the robust population transfer problem, provided that the
relative order between $\eps_1$ and $\eps_2$ satisfies some suitable constraint as both parameters go to zero and under some further assumptions on the range $[\al_0,\al_1]$. This is detailed in the following theorem. 
\begin{theorem}
\label{mainTh}
Assume that $v_0<0<v_1$ are such that 
$3(E+v_0) \ge E+v_1$. Fix $T>0$ and $u,\Delta :[0,T] \to \R$ smooth 
(e.g., $u\in C^2$ and $\Delta \in C^3$) such that
\be
\i $(u(0),\Delta'(0))=(0,2v_0)$ and $(u(T),\Delta'(T))=(0,2v_1)$;
\i $\forall s \in (0,T), u(s)>0$ and $\Delta''(s) \geq 0$.
\ee
 
Denote by $\psi^\al_{\eps_1,\eps_2}$ the solution of \eqref{1} with initial condition $\psi^\al_{\eps_1,\eps_2}(0)=(0,1)$ and control 
$w_{\eps_1,\eps_2}$ as in \eqref{control}. 
Then, for every $N_0 \in \N$,
%
for every compact interval $I \subseteq (v_0,v_1)$, there exist $C_ {N_0}>0$ and $\eta>0$ such that for every $\al \in I$  and every $(\eps_1,\eps_2)\in (0,\eta)^2$, 
\[|\psi^\al_{\eps_1,\eps_2}(\Tf)-(e^{i\theta},0)|< C_ {N_0} \max(\eps_2/\eps_1,\eps_1^{N_0-1}/\eps_2)\]
 for some $\theta\in\R$.
Moreover, the constant $C_{N_0}$ can be taken locally uniform with respect to the parameter $\delta>0$ appearing in \eqref{control}.
\end{theorem}
Roughly speaking, $\eps_2/\eps_1$ is  the AA error and $\eps_1^{N_0-1}/\eps_2$ the RWA error.
We define the fidelity of a pulse as the quantity $\inf_{\theta}|\psi^\al_{\eps_1,\eps_2}(\Tf)-(e^{i\theta},0)|$ (also denoted $|\langle \psi^\al_{\eps_1,\eps_2}(\frac{1}{\eps_1 \eps_2})| e_1\rangle|$). It is a natural measure of the transition rate induced by a pulse.
Thus, by playing on the integer $N_0$ and on the order of magnitude between $\eps_1$ and $\eps_2$, we can express the fidelity attained by the strategy above in terms of the duration of the pulse.
\begin{corollary}
Taking $\eps_1=\eps_2^{2/N_0}$ ($N_0\geq 3$) leads to an error of the order ${\cal T}^{\frac{2/N_0 -1}{1+2/N_0}}$, where ${\cal T}=1/(\eps_1\eps_2)$ is the duration of the pulse $w_{\eps_1,\eps_2}$.
\end{corollary}
\begin{remark}
    \label{remark:5}
\bi
\i As an example, one can 
apply Theorem~\ref{mainTh} with
 $T=1$, $\delta=1$, $\Delta(s)=\frac{v_0-v_1}{\pi}\sin(\pi s)+(v_0+v_1)s$ and $u(s)=1-\cos(2\pi s)$, $s\in [0,1]$. 
 More explicitly,
$$w_{\eps_1,\eps_2}(t)=2 \eps_1 (1-\cos(2\pi \eps_1 \eps_2 t))\cos\Big( 2Et+ \frac{(v_0-v_1)\sin(\pi \eps_1 \eps_2 t)}{\pi \eps_1\eps_2} +(v_0+v_1)t \Big). $$
 All the simulations in this paper use this pulse scheme and some compare to the complex-valued pulse
 $$w^{\mathrm{R}}_{\eps_1,\eps_2}(t)=\eps_1 (1-\cos(2\pi \eps_1 \eps_2 t))\exp{\Big( 2iEt+ i\frac{(v_0-v_1)\sin(\pi \eps_1 \eps_2 t)}{\pi \eps_1\eps_2} +i(v_0+v_1)t \Big)}. $$
\i By taking $N_0$ large, one can get, for each $\eta>0$, a fidelity close to one at order ${\cal T}^{-1+\eta}$, to compare with the standard $O({\cal T}^{-1})$ of the adiabatic theorem. 

\item 

The assumption $3(E+v_0) \geq E+v_1$  ensures non-overlapping of some characteristic frequencies (cf.~Lemma~\ref{lem:Omega}).
It could be replaced by the weaker one: 
$4E+3\Delta'-2\al>0$ for every $\al\in [\al_0,\al_1]$ 
and everywhere  in $[0,T]$. Nevertheless, asking this condition to be valid 
for every compact subinterval $[\tilde \al_0,\tilde\al_1]$ of $(v_0,v_1)$ 
is equivalent to the inequality 
$3(E+v_0) \geq E+v_1$. 

Numerical simulations suggest that 
the inequality $4E+3\Delta'-2\al>0$ is sharp in the following sense: if for a given $\al$, $4E+3\Delta'(s)-2\al<0$ for some $s\in [0,T]$, an inequality as in Theorem \ref{mainTh} seem not to hold.
As an illustration, in Figure~\ref{fig:range_al_withoutOmega} we observe that for $\al \geq 0$ (condition $4E+3\Delta'-2\al{\color{red}>}0$ not satisfied), the accuracy of the RWA is worse than for $\al <0$ (condition $4E+3\Delta'-2\al{\color{red}>}0$ satisfied).
\begin{figure}
    \begin{center}\includegraphics[width=0.7\columnwidth]{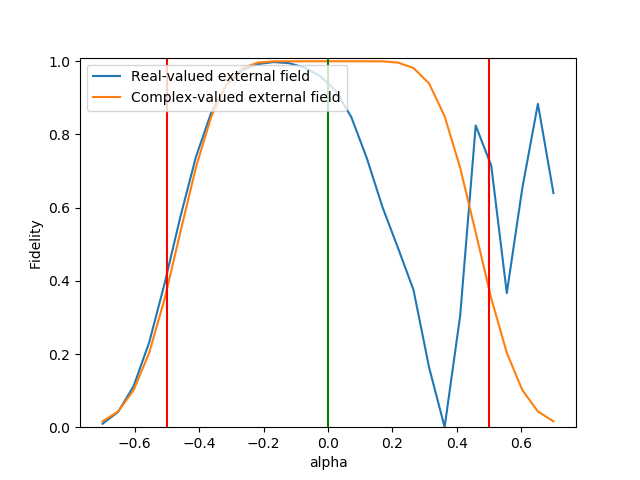}
    \caption{$E=0.75$, $\al=0.25$, $\eps_1=1$, $\eps_2=0.1$, $v_0=-0.5$, $v_1=0.5$. Assumption $4E+3\Delta'(s)-2\al<0$ is satisfied if and only if $\al <0$. \label{fig:range_al_withoutOmega}}
    \end{center}
    \end{figure}

\ei
\end{remark}

    \begin{remark}
    Many questions 
    concerning the combination of the RWA and AA
    remain 
   open. 
   In particular we do not know if a version of Theorem~\ref{mainTh} holds with $\eps_1$ fixed, small enough, and $\eps_2$ going to $0$.
    
    Concerning systems with higher number of levels (possibly infinite), we expect the techniques developed in this paper to work. Nevertheless, such an extension seems not trivial.
    \end{remark}

We postpone the proof of Theorem \ref{mainTh} to Section \ref{sec:proof}.
This proof is technical and is sketched in Sections \ref{subsec:first_change_of_variable} and \ref{subsec:idea_of_the_proof} (see also Remark \ref{rem:fin}).
\section{Application to the ensemble control problem}

We denote by $\sigma_x,\sigma_y,\sigma_z$ the Pauli matrices given by
\begin{align}
    \sigma_x= \begin{pmatrix}
        0&1\\1& 0
    \end{pmatrix},\quad \sigma_y=\begin{pmatrix}
        0&-i\\i& 0
    \end{pmatrix}, \quad \sigma_z= \begin{pmatrix}
        1&0\\0&-1
    \end{pmatrix},
\end{align}
and by ${\rm SU}_2$ the special unitary group of degree 2. We recall that its Lie algebra $\mathfrak{su}_2$ is generated by $i\sigma_x$, $i\sigma_y$, and $i\sigma_z$. 

There is a natural distance on ${\rm SU}_2$ induced by the norm of endomorphism on $\C^2$, which we denote $\|\cdot\|$.
Let $v_0<0<v_1$ and $0<\delta_m \leq \delta_M$.
Let $\mathcal{D}=[v_0,v_1]\times [\delta_m,\delta_M]$ be the compact set of the dispersion parameters
and endow $\mathcal{F}:=C^0(\mathcal{D},{\rm SU}_2)$ with the usual distance 
$d_{\F}(f,g):=\max_{d\in \D} \|f(d)-g(d)\|$.

Li and Khaneja proved in \cite{k2} the following ensemble operator controllability result.
\begin{theorem}[Li--Khaneja, 2009]
    \label{th:Li-Khaneja}
    For any control bound $K>0$, any target distribution $M_F\in \F$, and any $\eps >0$, there exist some $T>0$ and controls $u,v\in L^\infty([0,T],[-K,K])$ such that the solution of the equation
    \begin{align}
        \label{eq:2ctr} i\frac{d}{dt}M(\al,\delta,t)=((E+\al) \sigma_z + \delta u(t)  \sigma_x + \delta v(t)  \sigma_y)M(\al,\delta,t), \quad M(\alpha,\delta,0)=I_2,\quad \forall (\al,\delta)\in \mathcal{D}
    \end{align}
    satisfies $d_{\F}(M(\cdot ,\cdot ,T),M_F(\cdot ,\cdot ))<\eps$.
\end{theorem}
\begin{remark}
    \begin{itemize}
        \item The result was originally stated on ${\rm SO}_3$ for the Bloch sphere, the extension to ${\rm SU}_2$ stated in Theorem \ref{eq:2ctr} is straightforward.
        \item This is a very strong ensemble controllability result, as it tackles the controllability of the semigroups.
    \end{itemize}
    
\end{remark}
We extend here this result to the problem of a qubit driven by a single real control, thus replacing Equation \eqref{eq:2ctr} by
\begin{align}
    \label{eq:ctr} i\frac{d}{dt}M(\al,\delta,t)=((E+\al) \sigma_z + \delta u(t)  \sigma_x)M(\al,\delta,t), \quad M(\alpha,\delta,0)=I_2,\quad \forall (\al,\delta)\in \mathcal{D}.
\end{align}

One of the key ingredients of the proof of Theorem \ref{th:Li-Khaneja} is the existence of an adiabatic pulse inducing 
a propagator $U \in \F$ 
such that
$\max_{d\in \D} \min_{\theta \in [0,2 \pi]} \| U(d) (0,1)^T -(e^{i\theta},0)^T\|$ is arbitrarily small.

Theorem \ref{mainTh} ensures the following corollary.
\begin{corollary}
    \label{cor:adiab}
    Suppose that $3(E+v_0) > E+v_1$.
    Then, for any $K>0$ and any $\eps>0$, there exist $T>0$ and a control $u\in L^\infty([0,T],[-K,K])$ such that
    the solution of Equation \eqref{eq:ctr} satisfies $\max_{(\al,\delta)\in \D} \min_{\theta \in [0,2 \pi]} \|M(\al,\delta,T) (0,1)^T -(e^{i\theta},0)^T\|<\eps$.
\end{corollary}
Based on Corollary~\ref{cor:adiab}, we will prove 
the following result, which 
generalizes Theorem~\ref{th:Li-Khaneja} under the extra assumption on the $\al$-dispersion.
\begin{theorem}
    \label{th:ense_ctrl}
    Suppose that $3(E+v_0) > E+v_1$.
    Let $\epsilon >0$, $M_F\in \F$, and $K>0$. Then there exist $T>0$ and $u\in L^\infty([0,T],[-K,K])$ such that the solution of Equation \eqref{eq:ctr}
    satisfies $d_{\F}(M(\cdot ,\cdot ,T),M_F(\cdot ,\cdot ))<\eps$.
\end{theorem}

The proof, sketched below, is an adaptation of the arguments used in \cite{k2}.\\
Let $\mathcal{R}=\{M(\cdot,\cdot,T) \mid T>0 \text{, M is a solution of \eqref{eq:ctr} for some }u\in L^\infty([0,T],[-K,K])\}$. It is clear that $\mathcal{R}$ and its closure $\bar{\mathcal{R}}$ are semigroups of $\F$. We 
have to prove that $\bar{\mathcal{R}}= \F$.\\
\begin{lemma}\label{lem:sigmaz}
    For all $t$ in $\R$, $(\al,\delta) \mapsto e^{-t(E+\al)i \sigma_z}$ is in $\bar{\mathcal{R}}$.
\end{lemma}

\begin{proof}
    Using a null control in \eqref{eq:ctr} during a time $t\geq 0$, we get $(\al,\delta) \mapsto e^{-t(E+\al)i \sigma_z}$ belongs to $\mathcal{R}$.

 Let us prove that the result also holds for $t<0$. Set an arbitrary $\eps>0$. By Corollary~\ref{cor:adiab}, there exists $U^\eps\in \mathcal{R}$ such that 
 \[\max_{d\in \D} \min_{\theta \in [0,2 \pi]} \| U^\eps(d) (0,1)^T -(e^{i\theta},0)^T\|<\eps.\]
    Using Euler angle decomposition, there exist three functions $a_\eps,b_\eps,c_\eps$ from $\D$ to $[0,2\pi]$ (not necessarily continuous)
    such that $U^\eps(d)=e^{a_\eps(d)i \sigma_z}e^{b_\eps(d)i \sigma_x}e^{c_\eps(d)i \sigma_z}$ for every $d \in \D$.
    In particular, $\max_{d\in \D}|b_\eps(d)-\pi |$ is of order $\eps$, so that 
    $\sup_{d\in \D}\|U^\eps(d)-\tilde U^\eps(d) \|$ is also of order $\eps$, where    
    $\tilde U^\eps(d):= e^{a_\eps(d)i \sigma_z}e^{\pi i \sigma_x}e^{c_\eps(d)i \sigma_z}$. 
 For all $t>0$, we have that  $(\al,\delta) \mapsto e^{-t(E+\al)i \sigma_z}$ is in $\bar{\mathcal{R}}$, by using the control $u\equiv 0$.
Using the relation $e^{-\pi i \sigma_x}e^{r i\sigma_z} e^{\pi i \sigma_x}=e^{-r i\sigma_z}$, $r \in \R$, we deduce that 
    \begin{align*}
        \tilde U^\eps(d) e^{-t (E+\al) i \sigma_z}\tilde U^\eps(d)&=e^{a_\eps(d)i \sigma_z}e^{-\pi i \sigma_x}e^{c_\eps(d)i \sigma_z} e^{-t (E+\al) i \sigma_z} e^{a^\eps(d)i \sigma_z}e^{\pi i \sigma_x}e^{c_\eps(d)i \sigma_z} \\
        &= e^{t (E+\alpha)i \sigma_z},
    \end{align*}
 for every  $d=(\alpha,\delta)$ in $\D$. 
    This shows that $(\al,\delta) \mapsto e^{t(E+\al)i \sigma_z}$ is at distance of order $\eps$ from an element of ${\mathcal{R}}$,    
    concluding the proof.
\end{proof}
\begin{lemma}
    Let $u\in \R$. Then $(\al,\delta) \mapsto e^{u\delta i \sigma_x}$ is in $\bar{\mathcal{R}}$.
\end{lemma}
\begin{proof}
    Consider first the case $|u|\leq K$. Setting $V_n(\al,\delta)=e^{(-(E+\al) i \sigma_z + u\delta i \sigma_x)/n}$, one can easily check that the sequence $((\al,\delta)\mapsto (V_n(\al,\delta)e^{t(E+\al)i \sigma_z/n})^n)_{n\in \N}$ is in $\bar{\mathcal{R}}$ and converges to $(\al,\delta) \mapsto e^{u\delta i \sigma_x}$ in $\F$. This concludes the case $|u|\leq K$. 
    We deduce the general case using the fact that $\bar{\mathcal{R}}$ is a semigroup.
\end{proof}

Let 
\begin{equation}\label{defg}
\mathfrak{g}=\{X\in \mathcal{C}^0(\mathcal{D}, \mathfrak{su}_2)\mid \forall t\in \R, \;e^{tX}\in \bar{\mathcal{R}}\}.
\end{equation}
Thus $(\al,\delta) \mapsto \delta i \sigma_x$ and $ (\al,\delta) \mapsto (E+\al)i \sigma_z$ belong to $\mathfrak{g}$.
The space $\mathcal{C}^0(\mathcal{D}, \mathfrak{su}_2)$ has a natural addition, product, and Lie bracket. 
Moreover, it has the structure of Banach algebra using as norm the sup norm, denoted by $|\cdot|_\infty$. 
Before concluding the proof of Theorem~\ref{th:ense_ctrl}, let us to show that $\mathfrak{g}$ is a Lie algebra by proving that it is stable by addition and Lie bracket.
\begin{lemma} The set
    $\mathfrak{g}$ defined in \eqref{defg} is stable under addition and Lie brackets:
    \begin{align*}
        [\mathfrak{g},\mathfrak{g}]\subset \mathfrak{g} ,\quad \mathfrak{g}+\mathfrak{g} \subset \mathfrak{g}.
    \end{align*}
\end{lemma}
\begin{proof}
    Pick $X,Y\in \mathfrak{g}$. Let us first prove that $e^{t[X,Y]}\in \bar{\mathcal{R}}$ for every $t\in \R$. 
   To this purpose, consider $U(s)=e^{s X}e^{s Y}e^{-s X}e^{-s Y}$,  $s\in [0,1]$. Then there exists a constant $C>0$ depending only on $|X|_\infty$ and $|Y|_\infty$ such that $d_\F(U(\sqrt{s}),e^{s[X,Y]}) \leq C s^{3/2}$ for every $s\in [0,1]$.
   Using the fact that the application $x\mapsto x^n$ is $(n-1)$-Lipschitz on the unit ball of any Banach algebra, we get
    \[d_\F(U(\sqrt{s/n} )^{n},e^{s [X,Y]})\leq C (s/n)^{3/2} (n-1)\leq Cs^{3/2} n^{-1/2},\qquad s\in [0,1].\]
    As a consequence, $e^{s[X,Y]}\in \bar{\mathcal{R}}$ for every $s\in [0,1]$. Applying the same reasoning to $-X$ instead of  $X$, we get that $e^{s[X,Y]}\in \bar{\mathcal{R}}$ for every $s\in [-1,1]$.
    We conclude the proof of the stability under Lei bracket by using the semigroup structure of $\bar{\mathcal{R}}$.

    Concerning the stability under addition, set $V(s)=e^{sX}e^{sY}$ and notice that $V(s)\in \bar{\mathcal{R}}$  for every $s\in \R$. Noticing that $V(t/n)^n \xrightarrow[n\to \infty]{d_\F} e^{t(X+Y)}$, we deduce that $e^{t(X+Y)}\in \bar{\mathcal{R}}$ for every $t\in \R$.
\end{proof}
Denote by ${\rm ad}_X(Y)=[X,Y]$ the adjoint representation both in $\mathfrak{su}_2$ and in $\mathcal{C}^0(\mathcal{D}, \mathfrak{su}_2)$. We recall the Pauli matrices commutation laws 
$$[i\sigma_x,i\sigma_y]=-i\sigma_z, \quad [i\sigma_y,i\sigma_z]=-i\sigma_x, \quad [i\sigma_z,i\sigma_x]=-i\sigma_y.$$
After some straightforward computations, one gets
\begin{eqnarray*}
    {\rm ad}_{\delta i\sigma_x}^{2l}({\rm ad}_{(E+\al) i\sigma_z}^{2k+1}(\delta i\sigma_x))=&(-1)^{l+k}(E+\al) ^{2k+1}\delta^{2l+1} i\sigma_y,\\
    {\rm ad}_{\delta i\sigma_x}^{2l+1}({\rm ad}_{(E+\al) i\sigma_z}^{2k+1}(\delta i\sigma_x))=&(-1)^{l+k}(E+\al) ^{2k+1}\delta^{2l+2} i\sigma_z,\\
    {\rm ad}_{(E+\al) i\sigma_z}{\rm ad}_{\delta i\sigma_x}^{2l}({\rm ad}_{(E+\al) i\sigma_z}^{2k+1}(\delta i\sigma_x))=&(-1)^{l+k+1}(E+\al) ^{2k+2}\delta^{2l+1} i\sigma_x.
\end{eqnarray*}
Thus for any $n,m\in \N$, and any sequence $(b_{k,l})_{k,l}$ 
of real numbers, we have
\begin{eqnarray*}
\label{P0} \sum_{k=0}^m \sum_{l=0}^n b_{k,l}\delta^{2k+2}(E+\al)^{2l+1}i\sigma_x  \in  \mathfrak{g},\\
    \label{P1} \sum_{k=0}^m \sum_{l=0}^n c_{k,l}\delta^{2k+1}(E+\al)^{2l+1}i\sigma_y \in  \mathfrak{g},\\
    \label{P2} \sum_{k=0}^m \sum_{l=0}^n d_{k,l}\delta^{2k+1}(E+\al)^{2l+2}i\sigma_z \in  \mathfrak{g}.
\end{eqnarray*}
By the Stone--Weierstrass theorem, for any continuous function $f\in C(\D,\R)$ we can approximate $\frac{f(d)}{(E+\al) \delta^2}$ uniformly on $\mathcal{D}$  by polynomials of the form $\sum_{k=0}^m \sum_{l=0}^n b_{k,l}\delta^{2k}(E+\al)^{2l}$. This proves that $f(d)i\sigma_x \in \mathfrak{g}$. With a similar argument, we get $f(d)i\sigma_\star \in \mathfrak{g}$ for $\star=x,y,z$. 

Finally, let $\rho>0$ be such that $(a_1,a_2,a_3)\mapsto e^{a_1 i\sigma_x+ a_2i\sigma_y+ a_3 i\sigma_z}$ is a diffeomorphism between a neighborhood of $0$ in $\R^3$ and the ball of radius $\rho$ centered at $I_2$ in ${\rm SU}_2$. Then for every $M_F\in \F$
such that $d_\F(M_F,I_2)<\rho$ there exist $f_1,f_2,f_3 \in \mathcal{C}^0(\mathcal{D},\R)$ such that $M_F(d)=e^{f_1(d)i\sigma_x+f_2(d)i\sigma_y+ f_3(d)i\sigma_z}$. 
Thus $M_F \in \bar{\mathcal{R}}$.
Since $ \bar{\mathcal{R}}$ is a semigroup, we deduce that $\bar{\mathcal{R}}$ is both open and closed in $\F$, yielding that $\bar{\mathcal{R}}=\F$. 
This concludes the proof of Theorem \ref{th:ense_ctrl}.

\section{Proof of Theorem~\ref{mainTh}}
\label{sec:proof}
\subsection{A first change of variables}
\label{subsec:first_change_of_variable}
Let $w_{\eps_1,\eps_2}$ be as in \eqref{control}. In order to recast the equation 
\[i\frac{d}{dt} \psi=H\psi=((E+\al) \sigma_z+w_{\eps_1,\eps_2}\sigma_x)\psi\]
in the interaction frame, set 
\[\psi_{\rm I}(t) = e^{i(E+\alpha)\sigma_zt}  \psi(t), \quad E_1(t)=2\al t-\frac{\Delta(\eps_1\eps_2 t)}{\eps_1\eps_2},\quad E_2(t)=4E t+2\al t+\frac{\Delta(\eps_1\eps_2 t)}{\eps_1\eps_2},\]
and notice that
\[i\frac{d}{dt}\psi_{\rm I}=H_{\rm I}\psi_{\rm I},\]
where 
\begin{eqnarray*}
H_{\rm I}(t)&=&-(E+\al)\sigma_z + e^{i(E+\alpha)\sigma_zt}H(t)e^{-i(E+\alpha)\sigma_zt}\\
&=&\eps_1u(\eps_1\eps_2t)
\Big(
\ba{cc} 
0& e^{i E_1(t)
}+ e^{i E_2(t)
}\\
e^{-i E_1(t)
}+e^{-i E_2(t)
}&0\ea
\Big).
\end{eqnarray*}
We will assume without loss of generality that $ T=1 $. 
For $E\in \R$, define
\bqn
A(E)=
\Big(
\ba{cc} 
0& e^{iE}\\e^{-iE}&0\ea
\Big) ,&& B(E)=
\Big(
\ba{cc} 
0& -ie^{iE}\\ie^{-iE}&0\ea
\Big) .
\eqn
In terms of these new notations, 
we can rewrite  $H_{\rm I}(t)=\eps_1 u(\eps_1 \eps_2 t) A(E_1(t))+\eps_1 u(\eps_1 \eps_2 t) A(E_2(t))$, $t \in [0, \frac{1}{\eps_1 \eps_2}]$.

\begin{figure}[ht]
\begin{center}
\includegraphics[width=0.7\columnwidth]{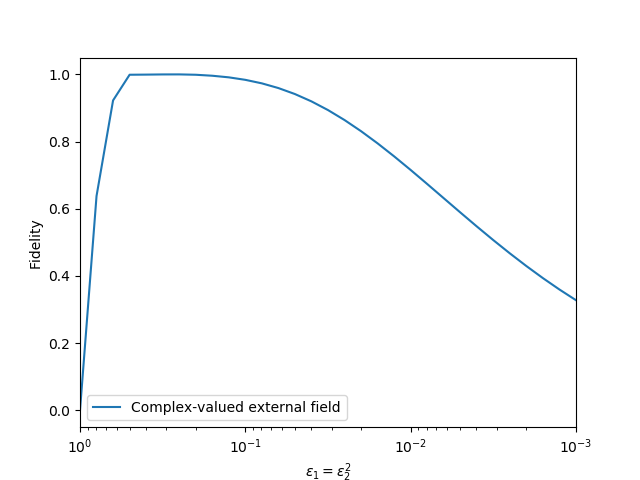}
\caption{Taking $v_0=-0.5$, $v_1=0.5$, $E=0$, and $\al=0$, we observe that the fidelity does not converge to $1$ 
as $(\eps_1,\eps_2)\to 0$
in the regime $\eps_1 \ll \eps_2$. 
The plot corresponds to the choice $\eps_1=\eps_2^2$. 
\label{fig:eps1/eps2}}
\end{center}
\end{figure}

In the usual first order RWA setting, one neglects the term containing the factor $
 A(E_2)$, which is highly oscillating compared to the first one. A standard method to justify this, is to use a change of variables close to the identity (see, e.g., \cite{arima:1904} and \cite{6b5219d6f6b54a8b930c33621289e040}). Inspired by this, we 
 introduce the notation
\begin{equation}
\label{eq:f12}
f_1(t)= \frac{d}{dt}E_1(t),\qquad 
f_2(t)= \frac{d}{dt}E_2(t),
\end{equation}
and we 
 apply the unitary change of variables
\bqn
\label{changeOfVariable}
\tilde \psi_{\rm I}(t) = \exp{\Big(i \eps_1\frac{u(\eps_1 \eps_2 t)}{f_2(t)}B(E_2(t))\Big) }\psi_{\rm I}(t).
\eqn
The dynamics of $\tilde \psi_{\rm I}$ are characterized by the Hamiltonian
\begin{align*}
    \tilde H_{\rm I}(t)={}& i \frac{d}{dt}\Big(\cos\Big(\eps_1 \frac{ u(\eps_1\eps_2t)}{f_2(t)}\Big)I+i \sin\Big(\eps_1 \frac{u(\eps_1\eps_2t)}{f_2(t)}\Big)B(E_2(t)) \Big)\\
    & \Big( \cos\Big(\eps_1 \frac{u(\eps_1\eps_2t)}{f_2(t)}\Big)I-i \sin\Big(\eps_1 \frac{u(\eps_1\eps_2t)}{f_2(t)}\Big)B(E_2(t))\Big)\\
    &+\Big(\cos\Big(\eps_1 \frac{ u(\eps_1\eps_2t)}{f_2(t)}\Big)I+i \sin\Big(\eps_1 \frac{u(\eps_1\eps_2t)}{f_2(t)}\Big)B(E_2(t))\Big) H_{\rm I}\\
    & \Big( \cos\Big(\eps_1 \frac{u(\eps_1\eps_2t)}{f_2(t)}\Big)I-i \sin\Big(\eps_1 \frac{u(\eps_1\eps_2t)}{f_2(t)}\Big)B(E_2(t))\Big).
\end{align*}
Notice that the first term can be rewritten as $-\eps_1 u(\eps_1\eps_2t)A(E_2(t))+O(\eps_1^2)$, so that $\tilde H_{\rm I}(t)=\eps_1 A(E_1(t)) + O(\eps_1^2+\eps_1^2\eps_2)$, where the notation $O(\cdot)$ is defined as follows.
\bdeff
\label{def:O}
Let $R$ be a \emph{$(\eps_1,\eps_2)$-parameterized function} in the following sense: 
for every $\epsilon_1,\epsilon_2>0$,  
$R_{\eps_1,\eps_2}$ is a real-valued function defined on the interval $[0,\frac{1}{\eps_1 \eps_2}]$.
We say that $R = O(g(\eps_1,\eps_2))$ with $g: \R_{+}^2 \to \R_+$ if there exist $\delta,C>0$ 
such that for every $(\eps_1,\eps_2)\in (0,\delta)^2$
and $t \in [0,\frac{1}{\eps_1 \eps_2}]$, we have $|R_{\eps_1,\eps_2}(t)| \leq C g(\eps_1, \eps_2)$.
\edeff
\begin{remark}
\label{introducingRWA}
\bi
\i We have $|\psi_{\rm I}- \tilde \psi_{\rm I}| = O(\eps_1)$. Moreover, from the hypotheses of Theorem~\ref{mainTh}, we have $u(0)=u(1)=0$, thus $\tilde \psi_{\rm I}(0)=\psi_{\rm I}(0)$ and $\tilde \psi_{\rm I}(\frac{1}{\eps_1\eps_2})=\psi_{\rm I}(\frac{1}{\eps_1\eps_2})$.
\i Let $\psi_{{\rm rwa}}$ 
be the solution of the Schr\"odinger equation with initial condition $\psi_{\rm I}(0)$ and Hamiltonian $\eps_1 u(\eps_1\eps_2 t) A(E_1(t))$. Then it turns out that $|\psi_{{\rm rwa}}(\frac{1}{\eps_1 \eps_2})-\tilde \psi_{\rm I}(\frac{1}{\eps_1 \eps_2})| = O(\eps_1/\eps_2)$ (see Lemma \ref{lemma:integ}). To prove convergence as $(\eps_1,\eps_2)\to 0$ in a suitable asymptotic regime, it would thus be enough to show that the dynamics of $\eps_1 u(\eps_1\eps_2t) A(E_1(t))$ induce a transition  
between $(0,1)$ and $(1,0)$ up to a phase, in the regime $\eps_1 \ll \eps_2$. Nevertheless this is not the case (recall that `standard' adiabatic theorem cannot be applied since $\eps_1$ is not fixed) as illustrated in Figure~\ref{fig:eps1/eps2}.
\ei
\end{remark}

\subsection{Idea of the proof}

\label{subsec:idea_of_the_proof}
We aim at providing correction terms to the Hamiltonian $\eps_1 u(\eps_1\eps_2 t) A(E_1(t))$, in order to improve the order of the averaging approximation. For this we will repeat a procedure similar to the one introduced in Equation~\eqref{changeOfVariable}. At each step 
the expression of the obtained effective Hamiltonian is more complicated  
 but provides a more accurate estimate of the final state. Then it will be possible to apply adiabatic theory to prove transition for the effective Hamiltonian. More precisely, we will prove the following theorem.
\bt
\label{mainAveraging}
Let $\al\in (v_0,v_1)$ and 
assume that $E+\al>0$ and $4E-3 \Delta'(s)>2\al$ for every $s\in [0,1]$. 
Then, for every $N_0\in \N$ there exists a Hamiltonian $H_{{\rm RWA}}$ of the form
\bqn
H_{{\rm RWA}}(t)=\eps_1 h_1(\eps_1\eps_2t) A(E_1(t))+\eps_1^2 h_2(\eps_1\eps_2t)B(E_2(t))+\eps_1^2 h_3(\eps_1\eps_2t)\sigma_z,
\eqn
with $h_1,h_2,h_3$ polynomials in $(\eps_1,\eps_2)$ with coefficients in $\mathcal{C}^{\infty}([0,1],\R)$,
such that the solution $\psi_{{\rm RWA}}$ of the Cauchy problem
\[
i\frac{d}{dt}\psi_{{\rm RWA}}=H_{{\rm RWA}} \psi_{{\rm RWA}}, \qquad \psi_{{\rm RWA}}(0)=\psi_{\rm I}(0),
\]
satisfies $|\psi_{{\rm RWA}}(\Tf)-\psi_{\rm I}(\Tf)| = O(\eps_1^2 \eps_2 + \eps_1^{N_0-1}/\eps_2)$.
 More precisely, there exist $h_{j,p,q} \in \mathcal{C}^{\infty}([0,1],\R)$, for $j=1,2,3$, $p=0,\ldots,N_0$, and $q=0,1$, such that \be
\i \label{mainAveragingpt1}$h_1= u +\sum_{p=1}^{N_0}\sum_{q=0}^1 \eps_1^p \eps_2^q h_{1,p,q}$ with $h_{1,p,0}(0)=h_{1,p,0}(1)=0$,
\i \label{mainAveragingpt1.5}$h_2= \sum_{p=0}^{N_0}\sum_{q=0}^1 \eps_1^p \eps_2^q h_{2,p,q}$ with $h_{2,p,0}(0)=h_{2,p,0}(1)=0$,
\i \label{mainAveragingpt2}$h_3= \sum_{p=0}^{N_0}\sum_{q=0}^1 \eps_1^p \eps_2^q h_{3,p,q}$ with $h_{3,p,0}(0)=h_{3,p,0}(1)=0$.
\ee
\et

After that, we will prove that $H_{{\rm RWA}}$ induces a transition between eigenstates with an error of order $O(\eps_2/\eps_1)$, which will be enough to prove Theorem~\ref{mainTh}.
\subsection{The rotating wave approximation}
\bigskip


\begin{definition}
Define the algebra $\mathcal{S}$ of \emph{slow functions} as the 
set of all $(\eps_1,\eps_2)$-parameterized functions $f$(in the sense of Definition~\ref{def:O}) 
such that for every $t\in [0,\Tf]$, $f_{\eps_1,\eps_2}(t)=g(\eps_1 \eps_2 t)$ for some smooth $g:[0,1]\to \R$ independent of $(\eps_1,\eps_2)$. The quantity $\sup_{t\in [0,\Tf]} |f_{\eps_1,\eps_2}(t)|$ is independent of $(\eps_1,\eps_2)$ and provides a norm, endowing $\mathcal{S}$ with the structure of Banach algebra.
\end{definition}
\begin{remark}
\bi 
\i The functions $f_1$ and $f_2$ defined in \eqref{eq:f12} 
are slow.
\i  $\mathcal{S}$ is isometric to the Banach algebra $\mathcal{C}^{\infty}([0,1],\R)$.
\i Given $f\in \mathcal{S}$, its $t$-derivative $\dot{f}$ defined by 
$\dot{f}_{\eps_1,\eps_2}(t)=\frac{d}{dt}{f}_{\eps_1,\eps_2}(t)$
is such that $\Tf \dot{f}\in \mathcal{S}$.
\ei 
 
\end{remark}
For every $j\in\Z$, let us  introduce the notations
\begin{align}
\Lambda_j={}&(j+1)E_1-jE_2, & \tilde \phi_j={}&jE_1-jE_2,\nonumber\\
\lambda_j={}&(j+1)f_1-jf_2, & \phi_j={}&jf_1-jf_2.\label{eq:frequences}
\end{align}
\bdeff
Define the set 
$$G=\{\pm Z(\Lambda_p), \pm \cos(\Phi_p) \sigma_z, \pm \sin(\Phi_p) \sigma_z  \mid Z \in \{A,B\},p \in \Z\}.$$ 
We say that an element of G is \emph{oscillating} if its associated integer $p$ is different from $0$.
\edeff
\bl
\label{stability_of_G}
G has the following stability properties:
\be
\i $\forall p \in \Z, \forall X\in G$, $\cos(\Phi_p)X$ and $\sin(\Phi_p)X$ are in ${\rm span}_\R G$;
\i $\forall X,Y \in G$, $i[X,Y] \in {\rm span}_\R G$;
\i $\forall X,Y \in G$, $XYX \in {\rm span}_\R G$.
\ee
\el
\bproof
The first point is a consequence of the fact that $\{\Phi_p \mid p \in \Z \}$ is a group for the addition. Thus, for every $p,q\in \Z$,
$$2\cos(\Lambda_p) \cos(\Phi_q)\sigma_z=\cos(\Phi_p+\Phi_q)\sigma_z+\cos(\Phi_p-\Phi_q)\sigma_z \in {\rm span}_\R G.$$
Moreover, $2\cos(\Phi_p)A(\Lambda_{q})=A(\Lambda_{q+p})+A(\Lambda_{q-p})\in {\rm span}_\R G$. The remaining cases can be checked similarly.

For the second point, for every $E,E',E''\in \R$,
\begin{align*}
i[A(E),A(E')]&=-2 \sin(E-E') \sigma_z, \\
i[A(E),\cos(E')\sigma_z)]&=2\cos(E')B(E)=B(E+E')+B(E-E'),\\
i [\cos(E') \sigma_z, \cos(E'') \sigma_z]&=0.
\end{align*}
Using the fact that, for every $p\in \Z$, $A(\Lambda_p-\pi/2)=B(\Lambda_p)$, we obtain that $i[A(\Lambda_p),G] \in {\rm span}_\R G$. Similar results can easily be obtained for $B(\Lambda_p)$, $\cos(\Phi_p)\sigma_z$, and $\sin(\Phi_p)\sigma_z$.

The last point relies on the relations
\begin{align*}
A(E)A(E')A(E)&=A(2E-E'),\\
\cos(E')A(E)\sigma_z A(E)&=-\cos(E')\sigma_z,\\
2\cos^2(E')\sigma_z A(E)\sigma_z&=A(E)+\frac{1}{2}(A(E+2E')+A(E-2E')),\\
2\cos^2(E')\cos(E'') \sigma_z^3&=\Big(\cos(E'')+\frac{1}{2}(\cos(2E'+E'')+\cos(2E'-E''))\Big)\sigma_z.
\end{align*}
\eproof

\bdeff
Define the vector space $\mathcal{E}$ as the set of entire series in $(\eps_1,\eps_2)$ 
with coefficients
in the set ${\rm span}_\mathcal{S} G $, i.e.,
\begin{align*}
\Big\{\sum_{j,k\geq 0}\eps_1^j \eps_2^k\sum_{g\in H_{j,k}} s_g g & \mid  H_{j,k}\subset G\text{ finite, } s_g \in \mathcal{S},\\
&\ \sum_{j,k\geq 0}\eps_1^j \eps_2^k\sum_{g\in H_{j,k}} |s_g| < \infty \text{ for $(\eps_1,\eps_2)$ small enough}\Big\}.
\end{align*}
\edeff
\subsubsection{The elimination procedure}

In order to generalize \eqref{changeOfVariable}, we introduce the operation of elimination of an oscillating term of a coefficient of $\mathcal{E}$.
\begin{definition}\label{pr}
Define the operation ${\rm Pr}:G\to G$ by the relations ${\rm Pr}(\pm A(\Lambda_p))=\pm B(\Lambda_p)$, ${\rm Pr}(\pm B(\Lambda_p))=\mp A(\Lambda_p)$, ${\rm Pr}(\pm \cos(\Phi_p)\sigma_z)=\pm \sin(\Phi_p)\sigma_z$, and ${\rm Pr}(\pm \sin(\Phi_p)\sigma_z)=\mp \cos(\Phi_p)\sigma_z$.
\end{definition}
\begin{definition}
\label{def:El}
Let $H \in \mathcal{E}$ and $Z(E)$ be an oscillating term of $G$ ($E=\Lambda_p$ if $Z\in \{A,B\}$ or $E=\Phi_p$ if $Z(E)\in \{ \cos(E)\sigma_z,\sin(E)\sigma_z\}$). Suppose that $f=\dot E$ (which is necessarily slow) is nowhere vanishing.
Fix $j\geq 1$, $k \geq 0$, $s\in \mathcal{S}$ and let $c=\eps_1^j \eps_2^k s$.
The operation of \emph{elimination of $cZ(E)$ from $H$} is defined as
\begin{align}
{\rm El}(c,Z(E))(H)={}&i\frac{d}{dt} \Big[ \exp \Big( i (c/f) {\rm Pr}(Z)(E) \Big) \Big] \exp \Big(-i(c /f) {\rm Pr}(Z)(E)\Big) \nonumber\\
& + \exp \Big( i (c/f) {\rm Pr}(Z)(E) \Big) H \exp \Big(-i(c /f) {\rm Pr}(Z)(E)\Big),\label{El}\\
\widetilde{{\rm El}}(c,Z(E)) (\psi)={}&\exp \Big(i c /f {\rm Pr}(Z)(E)\Big) \psi.\nonumber
\end{align}
\end{definition}
In fact, the elimination procedure is the generalization of the change of variables in  Equation~\eqref{changeOfVariable}. It transforms the Hamiltonian dynamics $i\frac{d}{dt}\psi=H \psi$ into the dynamics $i\frac{d}{dt}\eta={\rm El}(c,Z(E))(H) \eta$, where $\eta=\widetilde{{\rm El}}(c,Z(E))\psi$.
The term \emph{elimination} is motivated by the following lemma, stating that the procedure described above generates in the transformed Hamiltonian only terms of degree higher than   $\eps_1^j \eps_2^k$. 
\bl
\label{SR-strong}
Take $H,Z(E),j,k,c$ as in Definition~\ref{def:El}. Then
${\rm El}(c,Z(E))(H) \in \mathcal{E}$. Besides, if $H=O(\eps_1)$ then 
${\rm El}(c
,Z(E))(H+c
Z(E))=H+O(\eps_1^{j+1} \eps_{2}^{k})$.
\el
\bproof
First recall that for each matrix $M$ such that $M^2=I$ and each $c \in \R$, $\exp (icM)=\cos (c) I +i \sin (c) M$. As $A(E)^2=B(E)^2= \sigma_z^2 =I$, we can give an explicit expression for ${\rm El}(c,Z(E))(H)$.

Let us start from the case $Z(E)=A(\Lambda_p)$, for which we have
\begin{equation}
{\rm El}(c,A(\Lambda_p))(H)=J_1+J_2+J_3,\label{eq:ElA}
\end{equation}
where
\begin{align}
J_1&=i\frac{d}{dt}{(c/f)}\Big(-\sin(c/f)I+i\cos(c/f)B(\Lambda_p)\Big) \Big(\cos(c/f)I -i \sin(c/f) B(\Lambda_p)\Big)\nonumber\\
&=-\frac{d}{dt}{(c/f)}B(\Lambda_p),\nonumber\\
J_2&=- \sin(c/f)\frac{d}{dt}B(\Lambda_p)\Big(\cos(c/f)I -i \sin(c/f) B(\Lambda_p)\Big)\nonumber\\
&= -f \sin(c/f) A(\Lambda_p)\Big(\cos(c/f)I -i \sin(c/f) B(\Lambda_p)\Big),\nonumber\\
J_3&= \Big(\cos(c/f)I
 +i \sin(c/f) B(\Lambda_p)\Big) H \Big(\cos(c/f)I -i \sin(c/f) B(\Lambda_p)\Big).\nonumber
\end{align}
The term 
$J_1
$ 
is obviously an element of $\mathcal{E}$. Besides, $\cos(c/f)$ and $\sin(c/f)$ are entire series in $\eps_1,\eps_2$ with coefficients in $\mathcal{S}$. Thus,
\begin{equation*}
 J_2=-f \sin (c/f)\cos(c/f) A(\Lambda_p) -f \sin^2(c/f) \sigma_z \label{eq:secondterm}
\end{equation*}
 is also an element of $\mathcal{E}$.
 The last term to be considered is
\begin{equation*}
\label{eq:lastterm1}
J_3=\cos^2(c/f)H+\cos(c/f)\sin(c/f)i[B(\Lambda_p),H]+\sin^2(c/f)B(\Lambda_p)HB(\Lambda_p).
\end{equation*}
Thanks to Lemma~\ref{stability_of_G}, $J_3$ is then the sum of 
elements of $\mathcal{E}$.

Let us
now assume that $H=O(\eps_1)$ and 
 focus on the order of each term (in the case $Z(E)=A(\Lambda_p)$). 
 We notice that
 $J_1=O(\eps_1^{j+1} \eps_2^{k+1})$ as $\frac{d}{dt}(s/f)=O(\eps_1\eps_2)$ and $J_2=-c A(\Lambda_p) + O(\eps_1^{j+1}\eps_2^k)$. 
 Finally, $J_3=H+ (c/f)i[B(\Lambda_p),H]+O(\eps_1^{j+1} \eps_2^{k})$. As $H=O(\eps_1)$, we get $(c/f)i[B(\Lambda_p),H]=O(\eps_1^{j+1} \eps_2^{k})$. Thus
$${\rm El}(c
,Z(E))(H+c 
Z(E))=-c 
Z(E)+H+c 
Z(E)+O(\eps_1^{j+1} \eps_{2}^{k}).$$

The same computations as above work for the case $Z(E)=B(\Lambda_p)$.

In the case $Z(E)=\cos(\Phi_p)\sigma_z$ we have
\begin{equation}
{\rm El}(c,\cos(\Phi_p)\sigma_z)(H)=J_1+J_2+J_3,\label{Eq:Elc}
\end{equation}
where
\begin{align*}
J_1&=-\frac{d}{dt}{(c/f)}\sin(\Phi_p)\sigma_z,\\
J_2&=- c\cos(\Phi_p)\sigma_z, \\ 
J_3&= (\cos(c/f \sin(\Phi_p))I +i \sin(c/f\sin(\Phi_p)) \sigma_z) H 
(\cos(c/f \sin(\Phi_p))I -i \sin(c/f\sin(\Phi_p)) \sigma_z).
\end{align*}
Note that $\sin(c/f\sin(\Phi_p))$ and $\cos(c/f \sin(\Phi_p))$ can be developed as entire series in $\eps_1,\eps_2$ with coefficients in $\mathcal{S} \cos(\Phi_q)$ and $\mathcal{S} \sin(\Phi_q)$ for $q\in \Z$. Lemma~\ref{stability_of_G} ensures 
that ${\rm El}(c,\cos(\Phi_p)\sigma_z)(H)$ is an element of $\mathcal{E}$. The computations of the order of the terms when $H=O(\eps_1)$ are similar 
to those made above,
and one can apply the same reasoning to ${\rm El}(c,\sin(\Phi_p)\sigma_z)(H)$.
\eproof

A key assumption of Lemma~\ref{SR-strong} above is that $f$ is nowhere vanishing. The following result ensures that this is the case for all frequencies of the oscillating terms in $G$.

\bl\label{lem:Omega}
Let $j\in\Z$ be nonzero.
Then 
 the functions
$\lambda_j$ and $\phi_j$, defined as in \eqref{eq:frequences},
are nowhere vanishing in $[ 0,\Tf]$. 
\el
\begin{proof}
Let us first prove that
\begin{equation}
\label{Omega}
2f_1(t) <f_2(t),\qquad \forall t\in \Big[ 0,\Tf \Big],
\end{equation}
where we recall that $f_1$, $f_2$ are defined in \eqref{eq:f12}. 
Indeed, 
\[2f_1(t) -f_2(t)=2\alpha-4E-2\Delta'(\eps_1 \eps_2 t)-\Delta'(\eps_1 \eps_2 t)<2v_1-4E-6v_0,\]
where we used the inequality $\alpha<v_1$ and the fact that, according to the hypotheses of Theorem~\ref{mainTh},
 $\Delta'$ is increasing from $2v_0$ to $2v_1$. 
The inequality $2v_1-4E-6v_0=2(E+v_1)-6(E+v_0)\le 0$, corresponding to the assumption 
$3(E+v_0) \ge E+v_1$ of Theorem~\ref{mainTh},
 concludes the proof of \eqref{Omega}.
 
Moreover, 
\[f_2(t)=4E+2\alpha+\Delta'(t)\ge 4(E+v_0)\ge \frac{4(E+v_1)}3>0,\qquad \forall t\in \Big[ 0,\Tf \Big].\]

In particular, 
$f_1-f_2=-4E-2\Delta'\le -4(E+v_0)
<0$. 
This implies that $\phi_j$ never vanishes for $j\ne 0$.
Finally,
for $j>0$, $\lambda_j=(j+1)f_1-j f_2=(j-1)(f_1-f_2)+2f_1-f_2<0$, and, similarly, 
$\lambda_j =(j+1)(f_1-f_2)+f_2>0$ for $j<0$. 
%
\end{proof}

\subsubsection{Algorithm description}

We can now introduce an algorithm to simplify the Hamiltonian $H_{{\rm I}}$. The cleaning operation  ${\rm cl}_{\bar p}(p_0,q_0)$, with $p_0\le \bar p$, consists in eliminating from $H_{{\rm I}}$ all oscillating terms of degree $\eps_1^p\eps_2^q$ for 
\[\Big\{
    \begin{array}{ll}
        p \leq \bar p \\
        q<q_0
    \end{array}\qquad\mbox{ and }\qquad    \Big\{\begin{array}{ll}
        p\leq p_0 \\
        q=q_0
    \end{array}\]
     in lexicographic order on $(p,q)$.

The algorithm is constructed by induction, as follows:
\bi
\i ${\rm cl}_{p}(0,0)=H_{\rm I}$;
\i for $0\le p'<p$, ${\rm cl}_{p}(p'+1,q)$ is obtained from ${\rm cl}_{p}(p',q)$ by eliminating one by one all its oscillating terms of degree $(p'+1,q)$, using Lemma~\ref{SR-strong}; 
\i ${\rm cl}_{p}(0,q+1)={\rm cl}_{p}(p,q)$. Notice that, by construction, there is no term of degree $(0,q+1)$ in ${\rm cl}_{p}(p,q)$.
\ei
Associated with the transformed Hamiltonian ${\rm cl}_{p_0}(p,q)$, 
we define $\widetilde{{\rm cl}}_{p_0}(p,q)$ the variable obtained 
iteratively from $\psi_{\rm I}$ by applying, at every use of Lemma~\ref{SR-strong}, the corresponding transformation $\widetilde{{\rm El}}$.
\begin{remark} 
According to Lemma~\ref{SR-strong}, each elimination procedure produces only  terms of higher degree, thus the algorithm yielding ${\rm cl}_{p_0}(p,q)$ ends after a finite number of steps.
\end{remark}

When we apply the algorithm, we first deal with monomials of the type $\eps_1^p\eps_2^0$, $p\ge 1$. 
The following lemma provides a useful property concerning their corresponding coefficients.

\bl
\label{weakel-stable}
Define $\mathcal S_0=\{s\in \mathcal S\mid s(0)=s(\Tf)=0\}$. Given $p,p',q\in \N$ with $p'\le p$, consider the decomposition ${\rm cl}_{p}(p',q)=H_1+\eps_2 H_2$, where $H_1$ is an entire series in $\eps_1$ with coefficient in ${\rm span}_{\mathcal{S}}G$ and $H_2 \in \mathcal{E}$ ($H_1$ collects all the monomials of the type $\eps_1^n \eps_2^0$). Then the coefficients of $H_1$ are in ${\rm span}_{\mathcal{S}_0}G$.
\el
\bproof
Let us first consider the case $q=0$. Then $H=H_1+\eps_2 H_2$ and we want to eliminate an element $cZ(E)$ with $c=\eps_1^{p+1}s$ and $s\in \mathcal{S}_0$ using Formula $\eqref{El}$. Notice that
$$-\frac{d}{dt}{(c/f)}{\rm Pr}(Z)(E) \quad \text{and} \quad \exp \Big( i (c/f) {\rm Pr}(Z)(E) \Big) \eps_2 H_2 \exp \Big(-i(c /f) {\rm Pr}(Z)(E)\Big)$$
only consist of monomials of the type $\eps_1^n \eps_2^m$ with $m\geq 1$.

On the other hand, the terms $J_1$ and $J_2$ in Equations~\eqref{eq:ElA} and \eqref{Eq:Elc} (and the corresponding ones for $Z(E)=B(E)$ and $Z(E)=\sin(E)\sigma_z$) are clearly in ${\rm span}_{\mathcal{S}_0}G$.
Besides, the coefficients of $$\exp \Big( i (c/f) {\rm Pr}(Z)(E) \Big) H_1 \exp \Big(-i(c /f) {\rm Pr}(Z)(E)\Big)$$ also stay in ${\rm span}_{\mathcal{S}_0}G$, as $\mathcal S_0$ is a subalgebra.

In the case $q \not = 0$, the elimination of a term of degree $(p,q)$ with $q\geq 1$ does not impact the monomials of the type $\eps_1^n \eps_2^0$, according to Lemma~\ref{SR-strong}.
\eproof

Let $G_0$ be the set of non-oscillating elements of $G$.
\bl
Assume that \eqref{Omega} holds. Then we have 
\[{\rm cl}_{N_0}(N_0,1)=\eps_1 H_{N_0}+\eps_1^{{N_0}+1} H_{r,{N_0}}+\eps_1^2\eps_2 H'_{N_0}+ \eps_1^{{N_0}+1}\eps_2 H'_{N_0,r}+\eps_1^3\eps_2^2H''_r,\]
 where
\be
\i \label{l1} $H_{N_0}$ is a polynomial of degree $N_0-1$ in $\eps_1$ with coefficients in ${\rm span}_{\mathcal{S}_0} G_0$,
\i \label{l2} $H'_{N_0}$ is a polynomial of degree $N_0-2$ in $\eps_1$ with coefficients in ${\rm span}_{\mathcal{S}} G_0$,
\i \label{l3} $H_{r,N_0}$ is an entire series in $\eps_1$ with coefficients in ${\rm span}_{\mathcal{S}_0} G$,
\i \label{l4} $H'_{r,N_0}$ is an entire series in $\eps_1$ with coefficients in ${\rm span}_{\mathcal{S}} G$,
\i \label{l5} $H''_{r}$ is an entire series in $\eps_1,\eps_2$ with coefficients in ${\rm span}_{\mathcal{S}} G$.
\ee
\el
\bproof
Points 
\ref{l1} and~\ref{l3} follow from Lemma~\ref{weakel-stable}, while points~\ref{l2},~\ref{l4}, and~\ref{l5} follow from Lemma~\ref{SR-strong}.
\eproof

Noticing that, in particular, ${\rm cl}_{N_0}({N_0},1)=\eps_1 H_{N_0}+\eps_1^2\eps_2 H'_{N_0}+O(\eps_1^3\eps_2^2+\eps_1^{N_0})$, we introduce the truncation  $ H_{{\rm RWA}}=\eps_1 H_{N_0}+\eps_1^2\eps_2 H'_{N_0}$ of ${\rm cl}_{N_0}({N_0},1)$ and we denote by $\psi_{{\rm RWA}}$ the solution  of
\begin{equation}\label{eq:dynRWA}
i\frac{d}{dt} \psi_{{\rm RWA}}= H_{{\rm RWA}} \psi_{{\rm RWA}},\qquad \psi_{{\rm RWA}}(0)=\psi_{N_0}(0),
\end{equation}
where 
$\psi_{N_0}=\widetilde{{\rm cl}}_{N_0}({N_0},1)$.
Notice that, even if we are using the same notation $\psi_{{\rm RWA}}$, we are considering here a RWA 
of higher-order  than the one discussed in Remark~\ref{introducingRWA}.

\bl
\label{lemma:integ}
We have the following estimates:
\be
\i $|\psi_{N_0}(\Tf)-\psi_{\rm I}(\Tf)| =O(\eps_1^2\eps_2)$;
\i $|\psi_{N_0}(\Tf)- \psi_{{\rm RWA}}(\Tf)| =O(\eps_1^2 \eps_2 + \eps_1^{{N_0}-1}/\eps_2)$.
\ee
\el
\bproof
 By Lemma~\ref{weakel-stable}, all the changes of variable used for obtaining  ${\rm cl}_{N_0}(N_0,0)$ from $H_{\rm I}$ are of the form $\widetilde{{\rm El}}(c,Z(E)) (\psi)$ with $c=\eps_1^p s$, $s\in \mathcal{S}_0$. Thus $\psi_{\rm I}(0)= \widetilde{{\rm cl}}({N_0},0)(0)$ and $\psi_{\rm I}(\Tf)= \widetilde{{\rm cl}}_{N_0}({N_0},0)(\Tf)$. Such changes of variable preserve the state at the initial and final time.
After that we applied finitely many changes of variable of the form $\psi \mapsto \exp (i \eps_1^p \eps_2^q s Z(E))\psi $ with $p\geq 2$ and $q=1$. Thus 
\begin{equation}\label{eq:spaz}
\sup_{t\in [0,\Tf]} |\widetilde{{\rm cl}}_{N_0}({N_0},0)(t)-\psi_{N_0}(t)| =O(\eps_1 ^2 \eps_2),
\end{equation}
which concludes the proof of the first estimate.

Notice that 
\begin{align*}
\frac{d}{dt} |\psi_{N_0}- \psi_{{\rm RWA}}|^2={}&2 \operatorname{Re} i\langle \psi_{N_0}- \psi_{{\rm RWA}} |{\rm cl}_{N_0}({N_0},1)\psi_{N_0}-H_{{\rm RWA}} \psi_{{\rm RWA}}\rangle\\
={}&2 \operatorname{Re} (i \langle \psi_{N_0}- \psi_{{\rm RWA}} |
 H_{{\rm RWA}} 
(\psi_{N_0}- \psi_{{\rm RWA}})\rangle\\
&+i\langle \psi_{N_0}- \psi_{{\rm RWA}} |({\rm cl}_{N_0}({N_0},1)- H_{{\rm RWA}})\psi_{N_0}\rangle )\\
\leq& |\psi_{N_0}- \psi_{{\rm RWA}}|O(\eps_1^3\eps_2^2 + \eps_1^{N_0}).
\end{align*}
Thus, 
$$2\frac{d}{dt} |\psi_{N_0}- \psi_{{\rm RWA}}|\leq O(\eps_1^3\eps_2^2 + \eps_1^{N_0}),$$
and we conclude by integrating over $[0,\Tf]$.
\eproof
This concludes the proof of Theorem~\ref{mainAveraging}.

\subsection{Two scales adiabatic approximation}

The goal of this part is to prove the following lemma:
\bl
\label{adiab}
There exists $\delta>0$ such that the solution $\psi_{{\rm RWA}}$ of 
\eqref{eq:dynRWA}
satisfies $|\psi_{{\rm RWA}}(\Tf)-(e^{i\theta},0)|\leq M \eps_2/\eps_1$ for some $\theta \in \R$ (possibly depending on $\eps_1,\eps_2,\al$) for $(\eps_1,\eps_2)\in (0,\delta)^2$.
\el
With a slight abuse of notation, 
let us say in this section  that a  
$(\eps_1,\eps_2)$-parametric function $f$ is a $O(g(\eps_1,\eps_2))$ (respectively, a $\Omega(g(\eps_1,\eps_2))$) if  there exist $M,\delta>0$ such that
\bqn
\forall \eps_1,\eps_2 \in (0,\delta)^2, \forall s \in [0,1],|f_{\eps_1,\eps_2}(s)|\leq M g(\eps_1,\eps_2) \text{ (respectively, $|f_{\eps_1,\eps_2}(s)|\geq M g(\eps_1,\eps_2)$)}.
\eqn

Recall that 
$$H_{{\rm RWA}}(t)=\eps_1 h_1(\eps_1\eps_2t) A(E_1(t))+\eps_1^2 h_2(\eps_1\eps_2t)B(E_2(t))+\eps_1^2 h_3(\eps_1\eps_2t)\sigma_z,$$
with $h_1$, $h_2$, and $h_3$ given by Theorem~\ref{mainAveraging}.
We introduce the unitary change of variables $\psi_{\rm slow}(t)=U(t)\psi_{{\rm RWA}}(t)$ with
$$U(t)=\begin{pmatrix} e^{i(\alpha t -\frac{ \Delta(\epsilon_1 \epsilon_2 t)}{2\epsilon_1 \epsilon_2})} & 0 \\ 0 & e^{-i(\alpha t -\frac{ \Delta(\epsilon_1 \epsilon_2 t)}{2\epsilon_1 \epsilon_2})} \end{pmatrix}.$$
The notation $\psi_{\rm slow}$ is motivated by the fact that the Hamiltonian corresponding to its evolution is slow in the sense that it only depends on the \emph{slow variable} $s=\eps_1\eps_2 t$, also known as \emph{macroscopic} or \emph{reduced time}. More precisely, $i\frac{d}{dt}\psi_{\rm slow}(t)=H_{\rm slow}(\eps_1\eps_2 t)\psi_{\rm slow}(t)$, where
\bqn
H_{\rm slow}(s)&=&\eps_1 h_1(s) \sigma_x+\eps_1^2 h_2(s)\sigma_y+ \Big( \alpha-\frac{\Delta'(s)}{2} +\eps_1^2 h_3(s)\Big)\sigma_z.
\eqn
%
%
We cannot directly apply a `standard adiabatic theorem' to describe the evolution of $\psi_{\rm slow}$
because the adiabatic path 
depends on $(\eps_1,\eps_2)$.

The eigenvalues of $H_{\rm slow}(s)$ are 
\[\pm \omega_{\eps_1,\eps_2}(s)= \pm \sqrt{(\eps_1 h_1(s))^2+(\eps_1^2 h_2(s))^2+(\al-\Delta'(s)/2 +\eps_1^2 h_3(s))^2}, \qquad s\in [0,1].\]
Using a Taylor series development, we have $\omega_{\eps_1,\eps_2}=\Omega(\eps_1)$. Thus, for $(\eps_1,\eps_2)$ small enough, $\omega_{\eps_1,\eps_2}$ does not vanish.
As a consequence, we can introduce the spectral projector $P_{\eps_1,\eps_2}(s)$ of $H_{\rm slow}(s)$ on the negative eigenvalue.
Consider $\gamma_{\eps_1,\eps_2} :[0,1] \to S^2$ such that $H_{\rm slow}(s)=\omega_{\eps_1,\eps_2}(s) \gamma_{\eps_1,\eps_2}(s) \cdot \vec \sigma$ where  $\vec a \cdot \vec \sigma=a_1\sigma_x+a_2\sigma_y+a_3\sigma_z$. 
We want to approximate $P_{\eps_1,\eps_2}$ and its derivatives by the spectral projector on the negative eigenvalue for 
the simplified Hamiltonian $\tilde H_{\rm slow}= \eps_1 u \sigma_x + (\alpha-\Delta'/2)\sigma_z $ and its derivatives.

\bl
\label{lem:approxP}
Let $-\tilde \omega_{\eps_1}(s)$ be the negative eigenvalue  of the Hamiltonian $\tilde H_{\rm slow}(s)=\eps_1 u(s) \sigma_x + (\alpha-\Delta'(s)/2)\sigma_z $ and $\tilde P_{\eps_1}(s)$ be the spectral projector on $-\tilde \omega_{\eps_1}(s)$, $s\in [0,1]$. Then
\bi
\i $|P_{\eps_1,\eps_2}-\tilde P_{\eps_1}|=O(\eps_1)$,
\i $|\frac{d}{ds} P_{\eps_1,\eps_2}-\frac{d}{ds}\tilde P_{\eps_1}|=O(1)$,
\i $|\frac{d^2}{ds^2}P_{\eps_1,\eps_2}-\frac{d^2}{ds^2}\tilde P_{\eps_1}|=O(1/\eps_1)$.
\ei
\el
\bproof
First, remark that for every nonnegative 
integer $\nu$ \begin{equation}\label{eq:dir}
\frac{d^{\nu}}{ds^{\nu}}H_{\rm slow}(s)=\frac{d^{\nu}}{ds^{\nu}}\tilde H_{\rm slow}(s) + \frac{d^{\nu}}{ds^{\nu}}R(\eps_1,\eps_2,s),\end{equation}
where $\frac{d^{\nu}}{ds^{\nu}}R(\eps_1,\eps_2,s)=O(\eps_1^2)$.

For $H\in i\mathfrak{su}_2\setminus \{0\}$, 
define the orthogonal projector $P(H)$ as the projector on the negative eigenvalue of $H$. The map $P$ is $C^{\infty}$ and positively homogeneous of degree $0$ on $i\mathfrak{su}_2\setminus \{0\}$.

For every $r>0$, let $B_r$ be the Euclidean ball of center $0$ and radius $r$ in $\mathcal{M}_2(\C)$. 
Denote by $\mathcal{K}$ the compact set $i\mathfrak{su}_2 \cap \partial B_1$. The differential  $dP$ is positively homogeneous of degree $-1$, since
$$\forall H,h \in i\mathfrak{su}_2, dP_H(h)=dP_{\frac{H}{|H|}}\Big(\frac{h}{|H|}\Big).$$
As a consequence, for $H\in \mathfrak{su}_2 \setminus B_r$, $|dP_H|\leq \frac{\sup_{L \in \mathcal{K}} |dP_L|}{r}$.
Thus, there exists a universal constant $C>0$ such that $P$ is $\frac{C}{r}$-Lipschitz continuous on $i\mathfrak{su}_2\setminus B_r$.

Moreover, consider $r(\eps_1,\eps_2) := \inf_{s \in [0,1]}\omega_{\eps_1,\eps_2}(s)/2=\Omega(\eps_1)$. As $H_{\rm slow}-\Tilde H_{\rm slow}=O(\eps_1^2)$, for $\eps_1,\eps_2$ small enough we can assume that the segment $[H_{\rm slow}(s),\tilde H_{\rm slow}(s)]\cap B_{r(\eps_1,\eps_2)}$ is the empty set for every $s\in[0,1]$. Then, applying $P$ to the equality~(\ref{eq:dir}) for $\nu=0$, we obtain
\[ |P_{\eps_1,\eps_2}(s)-\tilde P_{\eps_1}(s)|\leq \frac{C}{r(\eps_1,\eps_2)}|R(\eps_1,\eps_2,s)| \leq M' \eps_1,\qquad  \forall s\in [0,1].\]

For the second point, we have $\frac{d}{ds} P_{\eps_1,\eps_2}(s)=dP_{H_{\rm slow}(s)}\Big( \frac{d}{ds}H_{\rm slow}(s)\Big)$. As $dP$ is positively homogeneous of degree $-1$, $d^2P$ is positively homogeneous of degree $-2$. Thus  $H \mapsto dP_H$ is $\frac{C'}{r^2}$-Lipschitz continuous on $i\mathfrak{su}_2\setminus B_r$ with $C'=\sup_{L \in \mathcal{K}} |d^2P_L|$. Thus, for $\eps_1,\eps_2$ small enough,
\[\Big| dP_{H_{\rm slow}}\Big( \frac{d}{ds}H_{\rm slow}\Big)-dP_{\tilde H_{\rm slow}}\Big( \frac{d}{ds} H_{\rm slow}\Big) \Big| \leq \frac{C'}{r(\eps_1,\eps_2)^2} \Big| H_{\rm slow}-\tilde H_{\rm slow} \Big|,
\]
and
\[\Big| dP_{\tilde H_{\rm slow}}\Big( \frac{d}{ds}H_{\rm slow}\Big)-dP_{\tilde H_{\rm slow}}\Big( \frac{d}{ds}\tilde H_{\rm slow}\Big)\Big| \leq \frac{C}{r(\eps_1,\eps_2)} \Big|\frac{d}{ds}H_{\rm slow}-\frac{d}{ds}\tilde H_{\rm slow}\Big|.
\]
Thus we get
\[dP_{H_{\rm slow}}\Big( \frac{d}{ds}H_{\rm slow}\Big)=dP_{\tilde H_{\rm slow}}\Big( \frac{d}{ds}\tilde H_{\rm slow}\Big) +O(1).\]
The third point is obtained by the same kind of argument.
\eproof

\begin{remark} The Hamiltonian $\tilde H_{\rm slow}(s) = \eps_1 u(s) \sigma_x + (\alpha-\Delta'(s)/2)\sigma_z $ is given by the first order RWA. The fact that $\eps_1$ appears in front of the pulse is obviously of utter importance for the estimation of the RWA error but also means that the `adiabatic path' is shrinking to the conical eigenvalue intersection. In fact, it is worse than just the shrinking of the spectral gap, as the derivative of the spectral projector is blowing up near the conical intersection (see Figures~\ref{chpdir1} and~\ref{chpdir2}). 

\begin{figure}[h!]
  \begin{center} 
  \begin{minipage}[t]{0.4\linewidth}
        \includegraphics[width=0.9\columnwidth]{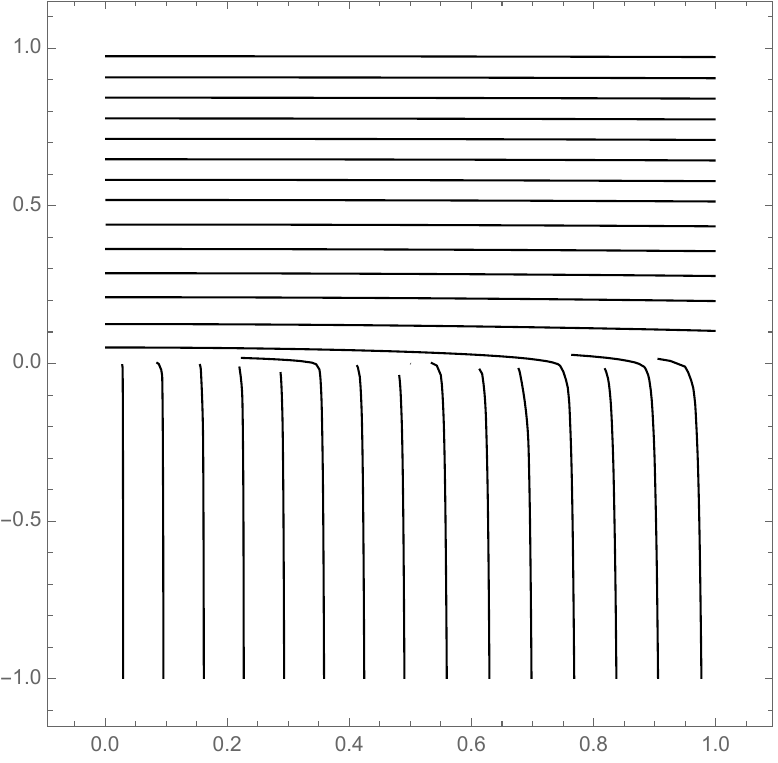}
        \caption{ Eigendirection corresponding to the negative eigenvalue of $\tilde H_{\rm slow}$ as a function of $(u,\Delta')\in \R^2$, for $\eps_1=0.01$ and $\al=0$. }
        \label{chpdir1}
    \end{minipage} 
\hspace{0.1cm}
    \begin{minipage}[t]{0.4\linewidth}
        \includegraphics[width=0.9\columnwidth]{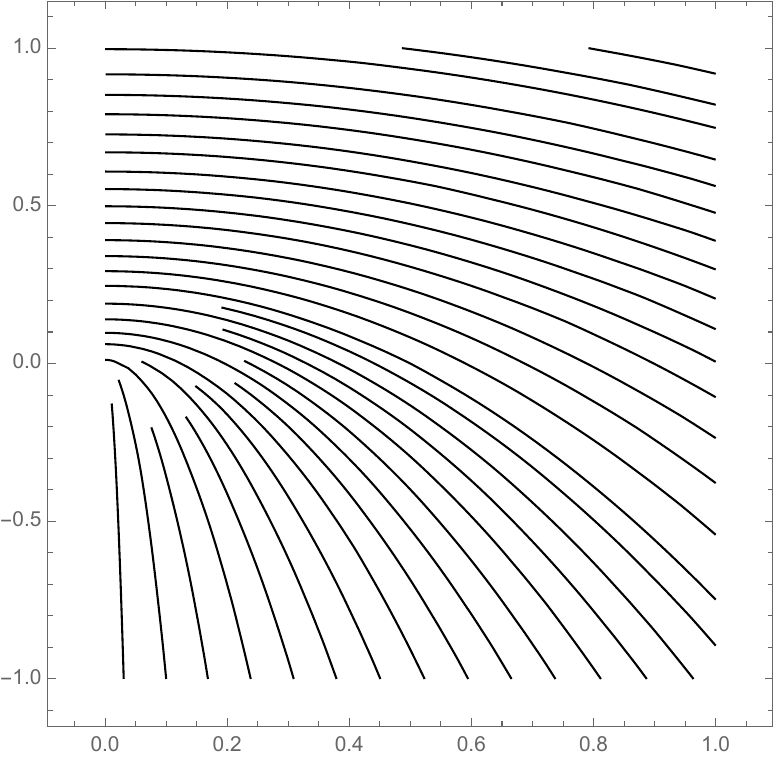}
        \caption{Eigendirection corresponding to the negative eigenvalue of $\tilde H_{\rm slow}$ as a function of $(u,\Delta')\in \R^2$, for $\eps_1=1$ and $\al=0$.}
        \label{chpdir2}
    \end{minipage}
    \end{center}
\end{figure}

\end{remark}

Define $\tilde \gamma_{\eps_1}(s)$, $s\in [0,1]$, by the relation $\tilde H_{\rm slow}(s)=\tilde\omega_{\eps_1}(s)\tilde \gamma_{\eps_1}(s) \cdot \vec \sigma$ and denote by  $(\tilde \theta,\tilde \phi)$ the spherical coordinates of $\tilde \gamma_{\eps_1}$.
Hence $X=\sin (\tilde \theta/2)e_1-e^{i \tilde \phi}\cos(\tilde \theta/2)e_2$ is an eigenvector of $\tilde H_{\rm slow}(s)$
associated with the negative eigenvalue and 
\bqn
\label{projector}
\tilde P_{\eps_1}= \Big( \ba{cc} \sin^2 (\tilde \theta/2) & -e^{-i\tilde \phi} \sin (\tilde \theta/2) \cos (\tilde \theta/2) \\
-e^{i\tilde \phi} \sin (\tilde \theta/2) \cos (\tilde \theta/2) & \cos^2 (\tilde \theta/2) \ea \Big).
\eqn

\bl
\label{lm:estimates}
Under the assumptions of Theorem~\ref{mainTh}, we have:
\be
\i \label{ll1} $|P_{\eps_1,\eps_2}(0)-P_{e_1}| = O( \eps_1^2 \eps_2)$ and $|P_{\eps_1,\eps_2}(1)-P_{e_2}| = O( \eps_1^2 \eps_2)$, where $P_{e_i}$ is the orthogonal projector on $\C e_i$;
\i \label{ll2} $\int_0^1 |(\frac{d}{ds} P_{\eps_1,\eps_2})(s)|^2 ds= O(1/\eps_1)$;
\i \label{ll3} $\int_0^1 |(\frac{d^2}{ds^2} P_{\eps_1,\eps_2})(s)| ds= O(1/ \eps_1)$;
\i \label{ll4}$ \int_0^1 |\frac{1}{\omega_{\eps_1,\eps_2}(s)^2} \frac{d}{ds}P_{\eps_1,\eps_2}(s)| | \frac{d}{ds} H_{\rm slow}(s)| ds= O(1/\eps_1^2)$.
\ee
\el
\bproof
Point~\ref{ll1} is a simple consequence of points~\ref{mainAveragingpt1},~\ref{mainAveragingpt1.5} and~\ref{mainAveragingpt2} in Theorem~\ref{mainAveraging}.

Concerning the other three points, thanks to Lemma~ \ref{lem:approxP} we are left to prove the corresponding estimates for $\tilde P_{\eps_1,\eps_2}$ and $\tilde H_{\rm slow}$.
%
We recall that $\tilde \theta(s)=\arccos \Big((\alpha-\Delta'(s)/2 )/ \tilde \omega_{\eps_1}(s)\Big)$. We can bound the transverse velocity of $ \tilde \gamma_{\eps_1}(s)$ by its total velocity 
\[\big|\tilde \omega_{\eps_1}(s) \frac{d}{ds} \tilde \theta(s) \big| \leq \sqrt{(\eps_1 \frac{d}{ds}u(s))^2+(\Delta''(s)/2)^2}=O(1),\]
 thus $\frac{d}{ds} \tilde \theta=O(1/\eps_1)$.

Using formula \eqref{projector}, it is clear that $|\frac{d}{ds}\tilde P_{\eps_1}(s)|\lesssim |\frac{d}{ds}\tilde \theta|$ and $|\frac{d^2}{ds^2}\tilde P_{\eps_1}(s)|\lesssim |\frac{d}{ds}\tilde \theta|^2 +|\frac{d^2}{ds^2}\tilde \theta|$, where $\lesssim$ stands for inequality up to an universal multiplicative constant. As $\Delta''\geq 0$, $\tilde \theta$ is increasing and
\begin{align*}
\int_0^1 \Big|\frac{d}{ds}\tilde \theta(s)\Big|^2 ds \leq \sup_{s\in [0,1]} \Big|\frac{d}{ds}\tilde \theta(s)\Big| \int_0^1 \Big|\frac{d}{ds}\tilde \theta(s)\Big| ds &\leq \pi \sup_{s\in [0,1]} \Big|\frac{d}{ds}\tilde \theta(s)\Big|= O(1/\eps_1).
\end{align*}
Moreover, bounding the transverse acceleration of $ \tilde \gamma_{\eps_1}(s)$ by its total acceleration, we have
$$\Big|2 \frac{d}{ds} \tilde \omega_{\eps_1}(s) \frac{d}{ds} \tilde \theta(s)- \tilde \omega_{\eps_1}(s) \frac{d^2}{ds^2} \tilde \theta(s) \Big| \leq \sqrt{(\eps_1 \frac{d^2}{ds^2}u(s))^2+(\Delta'''(s)/2)^2}=O(1).$$
As $\frac{d}{ds} \tilde \omega_{\eps_1}(s) \leq \sqrt{(\eps_1 \frac{d}{ds}u(s))^2+(\Delta''(s)/2)^2}=O(1)$, we have
$$ \int_0^1 \Big|\frac{d}{ds} \tilde \omega_{\eps_1}(s) \frac{d}{ds} \tilde \theta(s)\Big|=O(1),$$
leading to 
$$ \int_0^1 \Big|\frac{d^2}{ds^2} \tilde \omega_{\eps_1}(s)\Big|ds =O(1/\eps_1).$$

Concerning point 4, notice the integral $ \int_0^1 |\frac{1}{\omega_{\eps_1,\eps_2}(s)^2} \frac{d}{ds}P_{\eps_1,\eps_2}(s)| | \frac{d}{ds} H_{\rm slow}(s)| ds$
can be upper bounded, up to a multiplicative constant, by
$$\int_0^1 \Big( \frac{1}{\tilde \omega_{\eps_1}(s) ^2}\Big|\frac{d}{ds} \tilde \omega_{\eps_1}(s) \Big|  \Big|\frac{d}{ds} \tilde \theta(s)\Big| + \frac{1}{\tilde \omega_{\eps_1}(s)}\Big(\frac{d}{ds} \tilde \theta(s)\Big)^2 \Big)ds,$$ which is of order $1/\eps_1^2$.
\eproof

To conclude the proof of Lemma~\ref{adiab}, we 
deduce from  \cite[Corollary 2.3]{teufel-book} the adiabatic estimate
\begin{align*}
 \big|\psi^\al_{\eps_1,\eps_2}\big(\Tf\big)&-(e^{i\theta},0)\big|\leq  \eps_1 \eps_2\bigg[ \frac{|\frac{d}{ds} P_{\eps_1,\eps_2}(1)|}{\omega_{\eps_1,\eps_2}(1)}+\frac{|\frac{d}{ds} P_{\eps_1,\eps_2}(0)|}{\omega_{\eps_1,\eps_2}(0)}\\
 &+\int_0^1 \Big( \frac{2|\frac{d}{ds} P_{\eps_1,\eps_2}(s)|^2}{\omega_{\eps_1,\eps_2}(s)}+\frac{|\frac{d^2}{ds^2} P_{\eps_1,\eps_2}(s)|}{\omega_{\eps_1,\eps_2}(s)} + \frac{|\frac{d}{ds} P_{\eps_1,\eps_2}(s)| |\frac{d}{ds} H_{\rm slow}(s)|}{2\omega_{\eps_1,\eps_2}(s)^2} \Big) ds\bigg],
 \end{align*}
for some $\theta \in \R$.
Finally, Lemma~\ref{adiab} together with Theorem~\ref{mainAveraging} conclude the proof of Theorem~\ref{mainTh} for a given $\al$ and $\delta$. To get uniformity on the range of $\alpha$, notice that the algorithm does not depend on $\al$ elsewhere than in the expression of $E_1$ and $E_2$ (see \eqref{eq:f12}). For the adiabatic part, if we restrict $\al$ to a compact interval $I\subset (v_0,v_1)$, the estimates of Lemma~\ref{lm:estimates} can be taken uniform with respect to $\al$. The uniformity with respect to $\delta$ is straightforward.

\begin{remark}\label{rem:fin}
Now that we have detailed the whole proof, we want to stress some of its key points.
\begin{enumerate}
\i The changes of variables applied iteratively in order to eliminate the oscillating terms of the Hamiltonian 
induce 
a very small error (of order $\eps_2 \eps_1^2$) on the initial and 
the final state (Lemma~\ref{lemma:integ}), whereas the error is of order $\eps_1$ if one look at the entire trajectory.
\i 
The frequencies which appear during the algorithm 
are of very special type
 ($pf_1-pf_2$ and $(p+1)f_1-pf_2$ for $p$ integer) allowing us to perform as many changes of variables as we need and to give a simple condition implying that all such frequencies are nonzero.
\i \label{rem:oscR} Each change of variables yields a more complicated Hamiltonian. Fortunately, when we study the adiabatic dynamics of such an 
Hamiltonian, we can neglect all the terms except for those appearing in the first order RWA.
\i The first order RWA induces a population transfer in the limit $\eps_2 \ll \eps_1$.
\end{enumerate}
\end{remark}
\section{Numerical simulations}

We present in this section some numerical simulations illustrating the results stated in Theorem~\ref{mainTh}. 
In all simulations we use the chirp scheme 
presented in 
Remark~\ref{remark:5} with $E=1$, $v_0=-0.5$, and $v_1=0.5$.

\begin{figure}[H]
\begin{center}
\includegraphics[width=0.7\columnwidth]{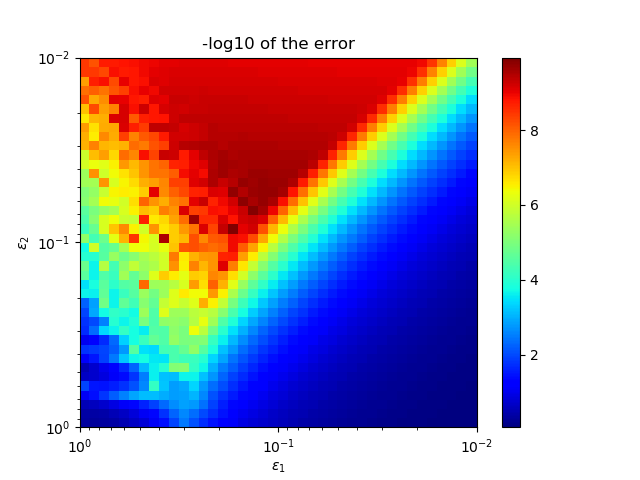}
\caption{Log of the distance from $\psi^0_{\eps_1,\eps_2}(\Tf)$ to the orbit of $(1,0)$.  \label{fig:2dplot}}
\end{center}
\end{figure}

Figure~\ref{fig:2dplot} shows the behavior of the distance from the target state as a function of $\eps_1,\eps_2$ represented in log scale. The AA error appears clearly, reflecting the fact that one needs $\eps_2 \ll \eps_1$ in order to have a fidelity close to 1. The figure also shows that the strategy has better performances than those anticipated theoretically in Theorem~\ref{mainTh}.

\begin{figure}[H]
\begin{center}
\includegraphics[width=0.7\columnwidth]{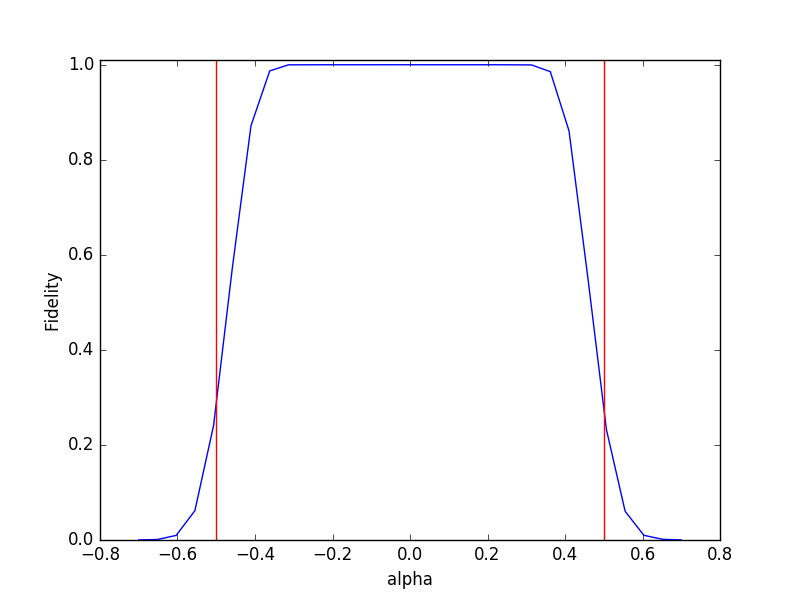}
\caption{Population transfer as a function of $\al$ for $E=1$, $\eps_1=0.5$ and $\eps_2=0.1$.  \label{fig:range_alpha}}
\end{center}
\end{figure}

Figure~\ref{fig:range_alpha} shows the fidelity as a function of $\alpha$, while  $\eps_1$ and $\eps_2$ (and hence ${\cal T}$) are fixed.

\begin{figure}[H]
\begin{center}
\includegraphics[width=0.7\columnwidth]{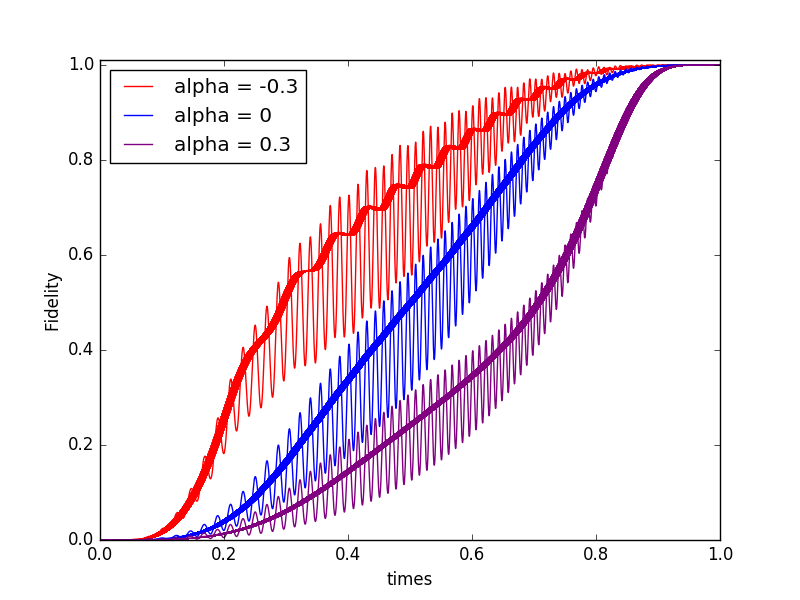}
\caption{$\eps_1=0.5$, $\eps_2=0.1$ and $\al=0$. In thick line are the
trajectories  corresponding to the equivalent 1st order RWA system.\label{fig:simu-time-alpha}}
\end{center}
\end{figure}

Figure~\ref{fig:simu-time-alpha} shows the fidelity as a function of the reduced time for three values of $\alpha$, while  $\eps_1$ and $\eps_2$ (and hence ${\cal T}$) are fixed. We clearly see that the RWA produces large oscillations (of magnitude of order $\eps_1$), which become much smaller at the endpoints, as described in Remark~\ref{rem:fin}, point~\ref{rem:oscR}.

\begin{figure}[H]
\begin{center}
\includegraphics[width=0.7\columnwidth]{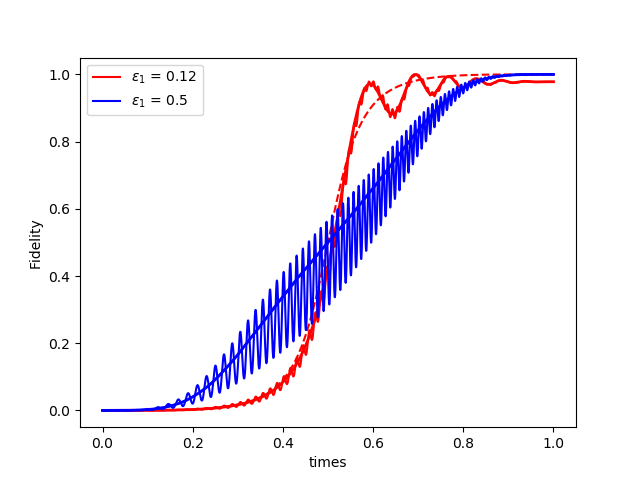}
\caption{$\eps_1 \eps_2=0.05$, $\al=0$. In thick line are the
trajectories corresponding to the equivalent 1st order RWA system and in dotted line the theoretical AA trajectories. \label{fig:simu-diff-eps1}}
\end{center}
\end{figure}

Finally, Figure~\ref{fig:simu-diff-eps1} illustrates the conflict between the AA and
RWA. At ${\cal T}=0.05$ fixed, for smaller $\eps_1$ we observe that the RWA is more accurate as
the thick line (1st order RWA) is closer to the highly oscillating one
(the trajectory $\psi^0_{\eps_1,\eps_2}$). Nevertheless as $\eps_1$ decreases, the ratio
$\eps_2/\eps_1$ increases and the AA becomes less accurate.
 
%

\section*{Acknowledgements}
The authors want to thank Steffen Glaser and Dominique Sugny for their very useful comments.

This work was supported by the ANR project SRGI ANR-15-CE40-0018, and by the ANR project Quaco ANR-17-CE40-0007-01.

\bibliographystyle{abbrv}
\bibliography{biblio-simultaneo}

\def\polhk#1{\setbox0=\hbox{#1}{\ooalign{\hidewidth
  \lower1.5ex\hbox{`}\hidewidth\crcr\unhbox0}}}
\begin{thebibliography}{10}

\bibitem{agrachevbook}
A.~A. Agrachev and Y.~L. Sachkov.
\newblock {\em Control theory from the geometric viewpoint}, volume~87 of {\em
  Encyclopaedia of Mathematical Sciences}.
\newblock Springer-Verlag, Berlin, 2004.
\newblock Control Theory and Optimization, II.

\bibitem{PhysRevA.75.063414}
S.~Ashhab, J.~R. Johansson, A.~M. Zagoskin, and F.~Nori.
\newblock Two-level systems driven by large-amplitude fields.
\newblock {\em Phys. Rev. A}, 75:063414, Jun 2007.

\bibitem{MR3874014}
N.~Augier, U.~Boscain, and M.~Sigalotti.
\newblock Adiabatic ensemble control of a continuum of quantum systems.
\newblock {\em SIAM J. Control Optim.}, 56(6):4045--4068, 2018.

\bibitem{augier:hal-02277852}
N.~Augier, U.~Boscain, and M.~Sigalotti.
\newblock {On the compatibility between the adiabatic and the rotating wave
  approximations in quantum control}.
\newblock In {\em {CDC 2019 - 58th Conference on Decision and Control}}, Nice,
  France, Dec. 2019.

\bibitem{semiconical}
N.~Augier, U.~Boscain, and M.~Sigalotti.
\newblock Semi-conical eigenvalue intersections and the ensemble
  controllability problem for quantum systems.
\newblock {\em Math. Control Relat. Fields}, 10(4):877--911, 2020.

\bibitem{RWAEXTENDED}
N.~Augier, U.~Boscain, and M.~Sigalotti.
\newblock Effective adiabatic control of a decoupled hamiltonian obtained by
  rotating wave approximation.
\newblock Preprint HAL : https://hal.inria.fr/hal-02562363, 2021.

\bibitem{BCR}
K.~Beauchard, J.-M. Coron, and P.~Rouchon.
\newblock Controllability issues for continuous-spectrum systems and ensemble
  controllability of {B}loch equations.
\newblock {\em Comm. Math. Phys.}, 296(2):525--557, 2010.

\bibitem{PhysRevA.82.022119}
X.~Cao, J.~Q. You, H.~Zheng, A.~G. Kofman, and F.~Nori.
\newblock Dynamics and quantum zeno effect for a qubit in either a low- or
  high-frequency bath beyond the rotating-wave approximation.
\newblock {\em Phys. Rev. A}, 82:022119, Aug 2010.

\bibitem{Garwood}
M.~Garwood and L.~DelaBarre.
\newblock The return of the frequency sweep: Designing adiabatic pulses for
  contemporary {NMR}.
\newblock {\em Journal of magnetic resonance (San Diego, Calif.: 1997)},
  153:155--77, 01 2002.

\bibitem{Jo2017}
H.~Jo, H.-g. Lee, S.~Guérin, and J.~Ahn.
\newblock Robust two-level system control by a detuned and chirped laser pulse.
\newblock {\em Physical Review A}, 96(3), Sep 2017.

\bibitem{rouchon-sarlette}
Z.~Leghtas, A.~Sarlette, and P.~Rouchon.
\newblock Adiabatic passage and ensemble control of quantum systems.
\newblock {\em Journal of Physics B}, 44(15), 2011.

\bibitem{k1}
J.-S. Li and N.~Khaneja.
\newblock Control of inhomogeneous quantum ensembles.
\newblock {\em Phys. Rev. A}, 73:030302, 2006.

\bibitem{k2}
J.-S. Li and N.~Khaneja.
\newblock Ensemble control of {B}loch equations.
\newblock {\em IEEE Trans. Automat. Control}, 54(3):528--536, 2009.

\bibitem{cdc-dasilva}
U.~A. Maciel~Neto, P.~S. Pereira~da Silva, K.~Beauchard, and P.~Rouchon.
\newblock ${H}^1$-control of an ensemble of half-spin systems replacing {R}abi
  pulses by adiabatic following.
\newblock In {\em Proceedings of the 58th IEEE Conference on Decision and
  Control}, 2019.

\bibitem{Malinovsky}
V.~Malinovsky and J.~Krause.
\newblock General theory of population transfer by adiabatic rapid passage with
  intense, chirped laser pulses.
\newblock {\em Eur. Phys. J. D}, 1450:147--155, 05 2001.

\bibitem{mittleman2013introduction}
M.~H. Mittleman.
\newblock {\em Introduction to the theory of laser-atom interactions}.
\newblock Springer Science \& Business Media, 2013.

\bibitem{arima:1904}
P.~Rouchon.
\newblock Quantum systems and control 1.
\newblock {\em Revue Africaine de la Recherche en Informatique et
  Mathématiques Appliquées}, Volume 9, Conference in Honor of Claude Lobry,
  2008.

\bibitem{6b5219d6f6b54a8b930c33621289e040}
J.~Sanders, F.~Verhulst, and J.~Murdock.
\newblock {\em Averaging methods in nonlinear dynamical systems}.
\newblock Number~59 in Applied Mathematical Sciences. Springer, 2007.

\bibitem{Scheuer_2014}
J.~Scheuer, X.~Kong, R.~S. Said, J.~Chen, A.~Kurz, L.~Marseglia, J.~Du, P.~R.
  Hemmer, S.~Montangero, T.~Calarco, B.~Naydenov, and F.~Jelezko.
\newblock Precise qubit control beyond the rotating wave approximation.
\newblock {\em New Journal of Physics}, 16(9):093022, sep 2014.

\bibitem{ccch11}
P.~Shapiro, Moshe;~Brumer.
\newblock {\em Quantum Control of Molecular Processes}.
\newblock John Wiley \& Sons, Ltd, 2012.

\bibitem{shore-article}
B.~Shore.
\newblock Coherent manipulations of atoms using laser light.
\newblock {\em Acta Physica Slovaca}, 58, 07 2008.

\bibitem{shore_2011}
B.~W. Shore.
\newblock {\em Manipulating Quantum Structures Using Laser Pulses}.
\newblock Cambridge University Press, 2011.

\bibitem{PhysRevLett.106.166801}
C.-M. Simon, T.~Belhadj, B.~Chatel, T.~Amand, P.~Renucci, A.~Lemaitre,
  O.~Krebs, P.~A. Dalgarno, R.~J. Warburton, X.~Marie, and B.~Urbaszek.
\newblock Robust quantum dot exciton generation via adiabatic passage with
  frequency-swept optical pulses.
\newblock {\em Phys. Rev. Lett.}, 106:166801, Apr 2011.

\bibitem{teufel-book}
S.~Teufel.
\newblock {\em Adiabatic perturbation theory in quantum dynamics}, volume 1821
  of {\em Lecture Notes in Mathematics}.
\newblock Springer-Verlag, Berlin, 2003.

\bibitem{PhysRevLett.106.233001}
B.~T. Torosov, S.~Gu\'erin, and N.~V. Vitanov.
\newblock High-fidelity adiabatic passage by composite sequences of chirped
  pulses.
\newblock {\em Phys. Rev. Lett.}, 106:233001, Jun 2011.

\bibitem{Vitanov-article}
N.~V. {Vitanov}, M.~{Fleischhauer}, B.~W. {Shore}, and K.~{Bergmann}.
\newblock {Coherent manipulation of atoms and molecules by sequential laser
  pulses}.
\newblock {\em Advances in Atomic Molecular and Optical Physics}, 46:55--190,
  Jan. 2001.

\bibitem{PhysRevLett.106.067401}
Y.~Wu, I.~M. Piper, M.~Ediger, P.~Brereton, E.~R. Schmidgall, P.~R. Eastham,
  M.~Hugues, M.~Hopkinson, and R.~T. Phillips.
\newblock Population inversion in a single {InGaAs} quantum dot using the
  method of adiabatic rapid passage.
\newblock {\em Phys. Rev. Lett.}, 106:067401, Feb 2011.

\end{thebibliography}

\end{document}